\documentclass[aps,prd,showpacs,amssymb,floatfix,nofootinbib,superscriptaddress]{revtex4-1}%\usepackage{setspace}
\usepackage{graphicx,amsmath,subfigure,psfrag}
\usepackage{multirow} 
\usepackage{xspace}
\usepackage{bm}
\usepackage{booktabs}
\usepackage{enumerate}
\usepackage{color}

\newcommand{\eeq}{\end{equation}}
\newcommand{\bea}{\begin{eqnarray}}

\def\ltsima{$\; \buildrel < \over \sim \;$}
\def\simlt{\lower.5ex\hbox{\ltsima}}
\def\gtsima{$\; \buildrel > \over \sim \;$}
\def\simgt{\lower.5ex\hbox{\gtsima}}

\def\lesssim{\mathrel{\hbox{\rlap{\hbox{\lower4pt\hbox{$\sim$}}}\hbox{$<$}}}}
\def\gtrsim{\mathrel{\hbox{\rlap{\hbox{\lower4pt\hbox{$\sim$}}}\hbox{$>$}}}}
\def\alt{\mathrel{\hbox{\rlap{\hbox{\lower4pt\hbox{$\sim$}}}\hbox{$<$}}}}
\def\agt{\mathrel{\hbox{\rlap{\hbox{\lower4pt\hbox{$\sim$}}}\hbox{$>$}}}}
\def\prd{{\it Phys. Rev.} D~}
\def\PRL{{\it Phys.Rev.} Lett~}
\def\apjl{{\it Astrophys. J.} Lett~}
\def\apj{{\it Astrophys. J.}}

\def\PRD{{\it Phys. Rev.} D~}
\def\CQG{{\it Class. Quantum Grav.}}

\def\pasj{{\it PASJ }}
\def\mnras{{\it MNRAS}} 
\def\aapr{{\it A\&ARv}}

\def\gta{\ifmmode {\mathbin{\lower 3pt\hbox   %> or of order
    {$\,\rlap{\raise 5pt\hbox{$\char'076$}}\mathchar"7218\,$}}}
    \else {${\mathbin{\lower 3pt\hbox
    {$\rlap{\raise 5pt\hbox{$\char'076$}}\mathchar"7218\,$}}}
    $}\fi}
\def\lta{\ifmmode {\,\mathbin{\lower 3pt\hbox   %< or of order
    {$\,\rlap{\raise 5pt\hbox{$\char'074$}}\mathchar"7218\,$}}}
    \else {${\mathbin{\lower 3pt\hbox
    {$\rlap{\raise 5pt\hbox{$\char'074$}}\mathchar"7218\,$}}}
    $}\fi}

\newcommand{\SU}{\affiliation{Department of Physics, Syracuse University, Syracuse, NY 13244, USA}}
\newcommand{\WVU}{\affiliation{Department of Physics, West Virginia University, White Hall, Morgantown, WV 26506, USA}}
\newcommand{\CU}{\affiliation{Institute of Astronomy, Madingley Road, CB3 0HA Cambridge, UK}}

\begin{document}

\title{Self-forced evolutions of an implicit rotating source: A natural framework to model  comparable and intermediate mass--ratio systems from inspiral through ringdown}
\author{E. A. Huerta}\email{elhuertaescudero@mail.wvu.edu}\WVU%
\author{Prayush Kumar}\email{pkumar@syr.edu}\SU%
\author{Jonathan R. Gair}\email{jgair@ast.cam.ac.uk}\CU%
\author{Sean T. McWilliams}\email{sean.mcwilliams@mail.wvu.edu}\WVU%

\date{\today}

\begin{abstract} 
We develop a waveform model to describe the inspiral, merger and ringdown of binary systems with comparable and intermediate mass-ratios. This model incorporates first-order conservative self-force corrections to the energy and angular momentum, which are valid in the strong--field regime~\cite{Akcay:2012}. We model the radiative part of the self-force by deriving second-order radiative corrections to the energy flux. These corrections are obtained by minimizing the phase discrepancy between our self-force model and the effective one body model~\cite{buho, Damour:2013} for a variety of mass-ratios. We show that our model performs substantially better than post-Newtonian approximants currently used to model neutron star-black hole mergers from early inspiral to the innermost stable circular orbit. In order to match the late inspiral evolution onto the plunge regime, we extend the ``transition phase'' developed by Ori and Thorne~\cite{amos} by including finite mass--ratio corrections and modelling the orbital phase evolution using an implicit rotating source~\cite{Baker:2008}. We explicitly show that the implicit rotating source approach provides a natural transition from late-time radiation to ringdown that is equivalent to ringdown waveform modelling based on a sum of quasinormal modes. 

\end{abstract}

\pacs{}
%\keywords{}

\maketitle

\section{Introduction}   
The black hole (BH) mass function in the local Universe is a strongly bi-modal distribution that identifies two main families: stellar-mass BHs with typical masses \(\sim 10M_{\odot}\) observed in Galactic X-ray binaries~\cite{McClintock:2006} and, more recently, in globular clusters~\cite{Morscher:2013}, and supermassive BHs with masses \(\gtrsim 10^{5} M_{\odot}\) observed to be present in most galactic nuclei~\cite{Merloni:2008, Fukugita:2004}.  However, a population of X-ray sources with luminosities in excess of \(10^{39}\, \rm{ erg}\, \rm{s}^{-1} \) has recently been observed, and {\it{Chandra}} and {\it{XMM-Newton}} spectral observations of these ultra-luminous X-ray sources (ULXs) revealed cool disc signatures that were consistent with the presence of intermediate mass BHs (IMBHs) with masses \(10^{2-4} M_{\odot}\)~\cite{Miller:2004,Miller:2004b,Miller:2006_BOOK}. Subsequent observations have shown that these ULXs have spectral and temporal signatures that are not consistent with the sub-Eddington accretion regime that is expected for IMBHs at typical ULX luminosities. Rather, these later studies suggest that many ULXs are powered by super-Eddington accretion onto \(\lesssim 100 M_{\odot}\) BH remnants.  Nevertheless, recent work by Swartz et al.~\cite{Swartz:2011} has demonstrated that as well as the high mass X-ray binaries that characterise most ULXs, there is a subpopulation of ULXs that seem to be powered by a separate physical mechanism. These objects have typical luminosities \(L\gtrsim 10^{41}\, \rm{ erg}\, \rm{s}^{-1} \), which cannot be explained by close to maximal radiation from super-Eddington accretion onto massive BHs formed in low metallicity regions~\cite{Zampieri:2009, Belczynski:2010, Ohsuga:2011}. Several hyper-luminous X-ray sources, including M82 X-1, ESO 243-49 HLX-1, Cartwheel N10 and CXO J122518.6+144545, present the best indirect evidence for the existence of IMBHs~\cite{Matsumoto:2001,Farrel:2009,Wolter:2010,Jonker:2010}.  In particular, the colocation of M82 X-1 with a massive, young stellar cluster, the features if its power spectrum, and some reported transitions between a hard state and a thermal dominant state, make this object a strong IMBH candidate~\cite{Portegies:2004,Strohmayer:2003,Kaaret:2007,Feng:2010}.  Recent searches of archival {\it{Chandra}} and {\it{XMM-Newton}} data sets have also uncovered two new hyper-luminous X-ray sources with luminosities in excess of \(10^{39}\, \rm{ erg}\, \rm{s}^{-1} \). These sources are the most promising IMBH candidates currently known, although the highest possible super-Eddington accretion rate onto the largest permitted BH remnant cannot yet be ruled out~\cite{Sutton:2012}.  This increasing body of observational evidence~\cite{Trenti:2006,Coleman:2004}, and the fact that the existence of IMBHs provides a compelling explanation for the initial seeding of supermassive BHs present in most galactic nuclei~\cite{Volonteri:2010,Schneider:2002,Yu:2002} has revived the quest for these elusive objects.  

Since hyper-luminous X-ray sources are rare, and our knowledge about their astrophysical properties is still limited, we may have to use a different means to search for IMBHs in order to improve our knowledge about the channels that lead to the formation of these objects, and to shed light on their astrophysical properties, such as mass and spin distributions~\cite{mandel}. In this paper we explore the use of observations in the gravitational wave (GW) spectrum to gain insight into the properties of IMBHs. 

\begin{figure*}[ht]
\centerline{
\includegraphics[height=0.35\textwidth,  clip]{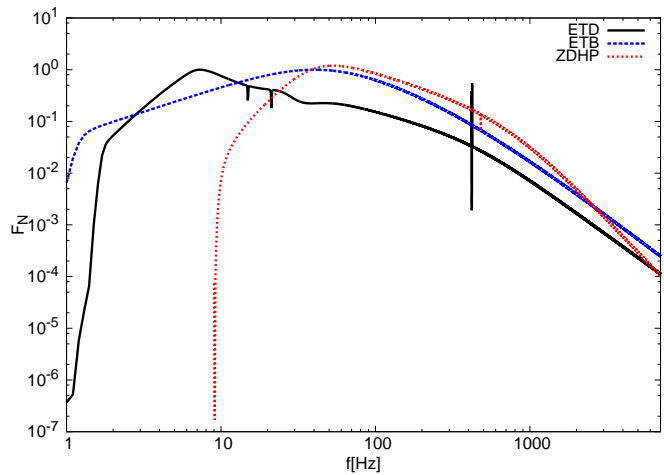}
}
\caption{The panel shows the expected sensitivity for two configurations of the Einstein Telescope (ET), namely, ETD (black), ETB (blue) and LIGO's Zero Detuned High Power (ZDHP) configuration (red). The vertical axis measures \(F_{\rm{normalized}} =  \left(f/f_{\rm{max}}\right)^{-7/6}\sqrt{S_n(f_{\rm{max}})/S_n(f)}\), where \(f_{\rm{max}}\) is the maximum of the corresponding power spectral density, \(S_n(f)\).
}
\label{ZDHP_promise}
\end{figure*}

The current upgrade of the Laser Interferometer Gravitational Wave Observatory (LIGO) and its international partners Virgo and Kagra~\cite{aLIGo, virgo,kagra}, will enable the detection of GWs from coalescences involving IMBHs with masses \(50 M_{\odot} \lesssim M \lesssim 500 M_{\odot}\), if these instruments achieve their target sensitivity down to the low-frequency cutoff at 10Hz (See Figure~\ref{ZDHP_promise})~\cite{ZDHP:2010}.  Advanced LIGO (aLIGO) and Advanced Virgo are expected to have greatest sensitivity in the 60Hz - 500Hz range, with a peak at \(\sim60\) Hz (see Figure~\ref{ZDHP_promise}). Proposed third generation detectors, such as the Einstein Telescope~\cite{Freise:2009}, which aim to extend the frequency range of ground-based detectors down to 1Hz, while also maintaining high frequency sensitivity up to 10kHz, will enhance our ability to search for GWs emitted by sources that involve BHs with masses between \(10^{2-4}M_{\odot}\)~\cite{etgair,Huerta:2011a,Huerta:2011b}.

A promising channel for detection of IMBHs is through the emission of gravitational radiation during the coalescence of stellar-mass compact remnants --- neutron stars (NSs) or BHs --- with IMBHs in core-collapsed globular clusters. This expectation is backed up by numerical simulations of globular clusters~\cite{Taniguchi:2000,Miller:2002,Mouri:2002a,Mouri:2002b,Gultekin:2004,Gultekin:2006,Oleary:2006,Oleary:2007}  which suggest that IMBHs could undergo several collisions with stellar-mass compact remnants during the lifetime of the cluster through a variety of mechanisms, including gravitational radiation, Kozai resonances and binary exchange processes.  As discussed in~\cite{man}, the most likely mechanism for the formation of binaries involving a stellar-mass compact remnant and an IMBH is hardening via three body interactions, with an expected detection rate of \(\sim 1-10\, {\rm{yr}}^{-1}\) with ground-based observatories~\cite{man,Abadie:2010}. 

The frequency of the dominant quadrupolar harmonic in the GWs emitted at the innermost stable circular orbit (ISCO) for a binary of non-spinning objects is 
\begin{equation}
f_{\rm{ISCO}}= 4.4 {\rm{kHz}} \left(\frac{M_{\odot}}{M}\right),
\label{fIMRIisco}
\end{equation}
\noindent so for a typical intermediate mass--ratio coalescence (IMRC) with total mass \(M\gtrsim 100 M_{\odot}\), advanced detectors will observe the late inspiral, merger and ringdown. However, the heaviest IMRCs, with masses $\gtrsim500M_\odot$, will coalesce out of band or at the low frequency limit of the bandwidth. Hence, merger and ringdown --- which intrinsically generate a signal-to-noise ratio (SNR) suppressed by a  factor of symmetric mass-ratio --- see Table~\ref{length} below ---  relative to the SNR generated during the inspiral phase--- will significantly contribute to the SNR of IMRCs over a considerable portion of the detectable mass-range~\cite{Smith:2013}. In order to get the most information from GW observations of IMRCs, it is therefore necessary to develop waveform models that incorporate the inspiral, merger and ringdown in a physically consistent way. In order to make progress in this direction, we previously developed a waveform model that combined results from Black Hole Perturbation Theory (BHPT) and post-Newtonian (PN) theory to  explore the information that could be obtained from observations of IMRCs with the EinsteinTelescope~\cite{Huerta:2011a,Huerta:2011b}. Although this model provided an important step in exploring the science that could be done with IMRC observations, the model was limited.

In~\cite{Huerta:2009,Huerta:2010,Huerta:EHE,Huerta:2012} we explored using the self-force formalism~\cite{SFB,LRP} to develop a waveform model with a robust description of the dynamical evolution of IMRCs during the inspiral phase. These were found to be effective when used to carry out matched-filter based searches for inspiral-only IMRCs~\cite{Smith:2013}, but searches for IMRCs in the advanced detector era will require waveform models that include not only the inspiral but also the merger and ringdown~\cite{Smith:2013}. The model described in~\cite{Huerta:2011a} included merger and ring down but without the self-force driven inspiral. It would be ideal to develop a consistent inspiral, merger, ringdown waveform model by comparison to numerical relativity simulations, as described in~\cite{buho,Damour:2013}. However, numerical relativity simulations for systems with mass-ratios \(q=m_1/m_2 \lesssim 1/10\) are very computationally expensive at present~\cite{Mroue:2013}. We can circumvent this problem by making use of recent breakthroughs in the self-force program that have shown that the conservative part of the self-force can reproduce, with good accuracy, results from numerical relativity simulations of comparable-mass binary systems~\cite{LeTiec:2012}.   Furthermore, the recent computation of the self-force inside the ISCO equips us to now develop models that better reproduce the true dynamics of black hole binaries in the strong field regime~\cite{Akcay:2012}.  

\begin{figure*}[ht]
\centerline{
\includegraphics[height=0.35\textwidth,  clip]{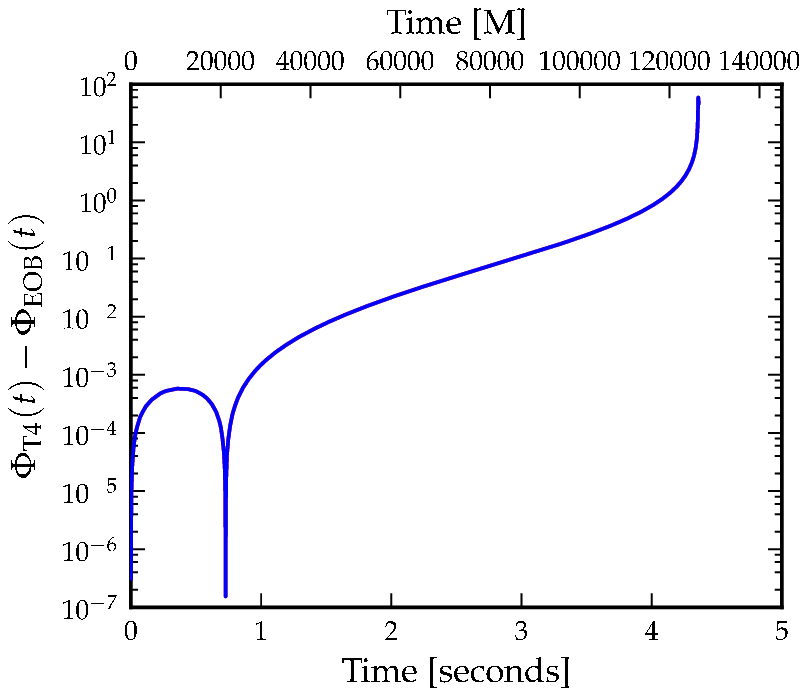}
\includegraphics[height=0.35\textwidth,  clip]{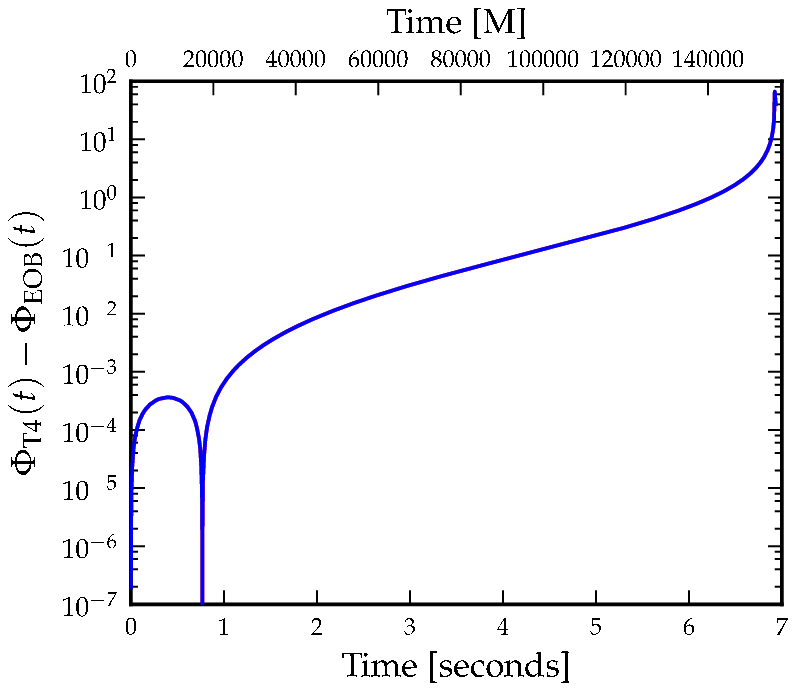}
}
\centerline{
\includegraphics[height=0.35\textwidth,  clip]{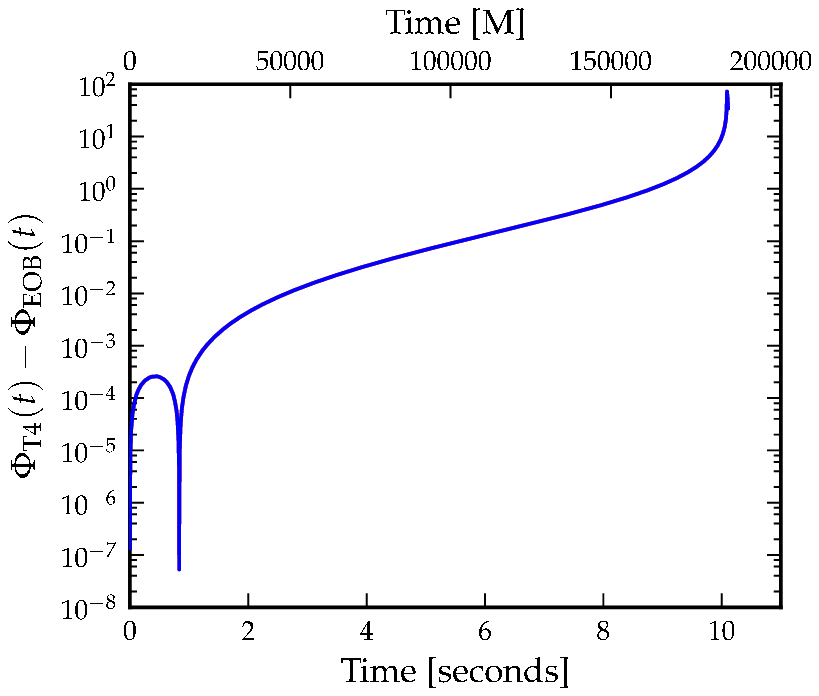}
\includegraphics[height=0.35\textwidth,  clip]{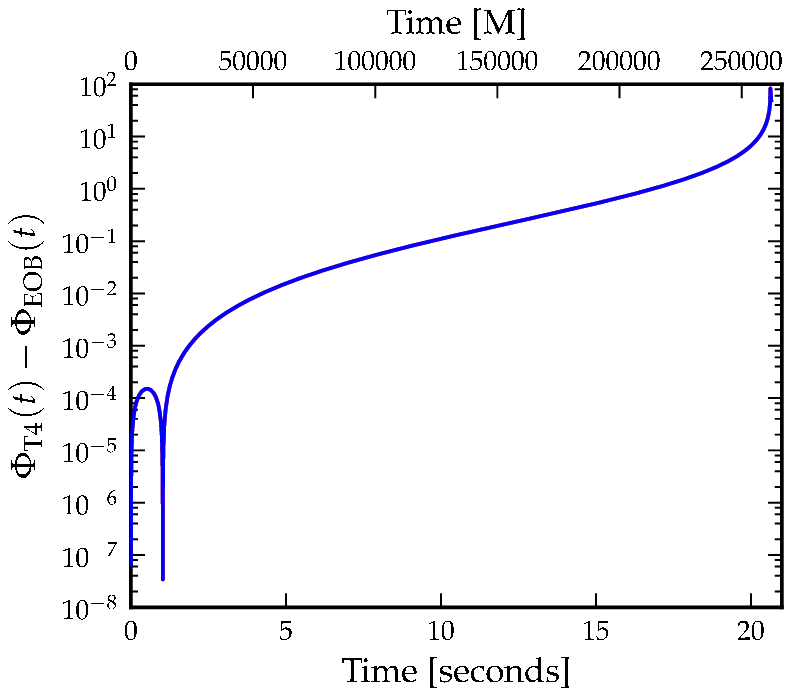}
}
\caption{The phase discrepancy in radians between the PN approximant TaylorT4, and the Effective One Body model, shown as a function of time from \(r=30M\) to the point when the TaylorT4 model reaches the ISCO. The systems have mass-ratio, $q$, total mass, \(M\), and final phase discrepancy, $\Delta\Phi$: $(q, M, \Delta\Phi) = (1/6, 7M_{\odot},21.5\,{\rm rads})$ (top-left), $(1/8, 9M_{\odot},30.2\,{\rm rads})$ (top-right), $(1/10, 11M_{\odot},70.1\,{\rm rads})$ (bottom-left) and $(1/15, 16M_{\odot},83.2\,{\rm rads})$ (bottom-right) respectively.}
\label{pn_approx}
\end{figure*}

In addition to IMRCs, which may be detected by second generation detectors, and with more likelihood by third generation detectors, we have explored the suitability of using self-forced evolutions to model the mergers of systems involving NSs and stellar mass BHs~\cite{LVCT}. Since NSBH mergers are promising GW sources for second generation detectors, we need to develop accurate and computationally inexpensive templates appropriate for these events. Current efforts to model NSBH systems have generally used PN approximants, evaluated to the highest PN order available, but these approximants are not sufficiently reliable to model these events~\cite{pnbuo,Nitz:2013mxa}. In Figure~\ref{pn_approx}, we show the phase difference between the PN approximant TaylorT4~\cite{TaylorT4Origin} and the EOB model introduced in~\cite{buho, Damour:2013}. This exhibits a substantial discrepancy near the ISCO. These considerations have impelled us to develop a different approach to model events with the typical mass-ratios expected for NSBH binaries.

In this paper we combine recent developments in the self-force program, in PN theory and in numerical relativity to develop a model that describes the inspiral, merger and ringdown of IMRCs and comparable mass-ratio systems that could be detected by second and third generation ground-based GW detectors. In Section~\ref{one} we discuss the modelling of the conservative part of the self-force. We show that using the linear in mass-ratio self force results for binaries with mass-ratios \(q\gtrsim1/6\) gives a system without an ISCO. We then discuss the implications of this result for the modelling of comparable and intermediate mass--ratio binaries.  Thereafter, we describe the approach we use to model the radiative part of the self-force for the inspiral evolution. Having constructed the inspiral part of the self-forced waveform model, we extend the transition scheme of  Ori and Thorne~\cite{ori} by including finite mass-ratio corrections, and modelling the orbital phase evolution using the implicit rotating source (IRS) model. We adopt this description for the late-time radiation in order to provide a smooth progression from late inspiral to rindgown.  We show that this approach provides the correct orbital frequency evolution in the vicinity of the light-ring. Finally, we construct the ringdown waveform using both a sum of quasinormal modes and the late-time radiation waveform evolution predicted by the IRS model, and show their equivalence. Section~\ref{conclu} presents a summary of our findings and future directions of work.

\section{Modeling}
\label{one}

\subsection{Nomenclature}
Throughout this paper we will use units with \(G=c=1\), unless otherwise stated. We will consider BH  binaries on circular orbits with component masses \(m_1, m_2\), such that \(m_1 < m_2\). We assume that the binary components are non spinning. We will use several combinations of the masses \(m_{1\, ,2}\) in the following Sections, which are summarized in Table~\ref{length}.

\begin{table}[thb]
\centering
\begin{tabular}{|c| c| }
\hline
\multicolumn{2}{|c|}{Binary masses}  \\\cline{1-2} 
\(m_1\) & mass of inspiralling compact object  \\ [0.7ex] 
\(m_2 \) & mass of central compact object  \\ [0.7ex]  
\(M=m_1+m_2 \) & total mass of binary system \\ [0.7ex]  
\(q=\frac{m_1}{m_2} \) & mass--ratio \\ [0.9ex]  
\(\mu=\frac{m_1 m_2}{m_1 + m_2} \) & reduced mass \\ [0.9ex]  
\(\eta= \frac{\mu}{M}\) & symmetric mass--ratio \\ [0.9ex]  
\hline
\end{tabular}
\caption{The table summarizes the nomenclature we will use throughout our analysis.}
\label{length}
\end{table}

Having defined the variables to be used in the subsequent sections, we shall now describe the construction of the self-forced waveform model. The model consists of four building blocks --- the inspiral, the transition, the plunge and the ringdown phases. The next section describes the inspiral evolution. 

\subsection{Inspiral evolution}

We model the inspiral phase evolution in the context of the Effective One Body  (EOB) formalism~\cite{EOB:Damour}, i.e., we consider the scenario in which the dynamics of a binary system is mapped onto the motion of a test particle in a time-independent and spherically symmetric Schwarzschild space-time with total mass \(M\):
\begin{equation} 
\mathrm{d}s^{2}_{\rm{EOB}} = -A(r)\mathrm{d}t^2 + B(r)\mathrm{d}t^2 + r^2\mathrm{d}\Omega^2\,,
\label{metricEOB}
\end{equation} 
\noindent where the potentials \(A, \, B\) are known to 3PN order~\cite{Buonanno:1999,Damour:2000}. In the test-mass particle limit \(\eta\rightarrow 0\), these potentials recover the Schwarzschild results, namely:
\begin{equation}
A(u, \eta\rightarrow 0) = B^{-1}(u,  \eta\rightarrow 0)= 1-2\,u,\quad {\rm{with}} \quad u=\frac{M}{r}.
\label{limitEOB}
\end{equation}
\noindent In the EOB formalism, the orbital frequency evolution can be derived from a Hamiltonian, \(H_{\rm{EOB}}\)~\cite{EOB:Damour}, given by:
\begin{equation} 
H_{\rm{EOB}} = M\sqrt{1+2\,\eta\left(H_{\rm{eff}} -1\right)},
\label{EOBH}
\end{equation}
\noindent using the general Hamiltonian equation:
\begin{equation}
\frac{d \phi}{d \mathrm{t} } =  M\Omega = \frac{\partial H_{\rm{EOB}}}{\partial L} = \frac{u^2\,L(x)\,A(u)}{H(u)\,H_{\rm{eff}}(u)},
\label{new_phase}
\end{equation}
\noindent where 
\begin{eqnarray}
H_{\rm{eff}}(u) &=& \frac{A(u)}{\sqrt{\tilde{A}(u)}}\,, \qquad  \tilde{A}(u)= A(u) + \frac{1}{2}\,u\,A'(u),  \\\nonumber  &&{\rm{and}}  \,\, \quad H(u)=\sqrt{1+2\,\eta\left(H_{\rm{eff}} -1\right)}.\\\nonumber
\label{params_for_new_phase}
\end{eqnarray}
\noindent Recent work in the self-force formalism has enabled the derivation of gravitational self-force corrections to the EOB potential \(A(u)\rightarrow  1-2u+\eta\, a(u) + {\cal{O}}(\eta^2)\)~\cite{barus}. Deriving this gravitational self-force contribution,  \(a(u)\), is equivalent to including all PN corrections to the EOB potential \(A(u)\) at linear order in \(\eta\). We shall now briefly describe the construction of the 
 gravitational self-force contribution \(a(u)\), emphasizing the fact that this contribution encodes information about the strong-field regime of the gravitational field. 

As shown by Detweiler and Whiting~\cite{Detweiler:2003}, the gravitational self-force corrected worldline can be interpreted as a  geodesic in a smooth perturbed spacetime with metric
 \begin{equation}
 g_{\alpha \beta} =  g^{0}_{\alpha \beta}(m_2) + h^{R}_{\alpha \beta},
 \label{detint}
 \end{equation}
 \noindent where the regularized \(R\) field is a smooth perturbation associated with \(m_1\). Detweiler proposed a gauge invariant relation to handle the conservative effect of the gravitational self-force in circular motion~\cite{Detweiler:2008,Detweiler:2009}:
 \begin{equation}
z_1(\Omega)= \sqrt{1-3x}\left(1- \frac{1}{2} h^{R,\,F}_{uu} + q \frac{x}{1-3x}\right),
\label{detinv}
\end{equation}
\noindent where  \(x\) is the gauge-invariant dimensionless frequency parameter given by \(x=\left(M \Omega\right)^{2/3}\), \(h^{R,\,G}_{uu}\) is a double contraction of the regularised metric perturbation with the four-velocity, \(u^{\mu}\), \( h^{R,\, G}_{uu} = h^{R,\,G}_{\mu\nu}u^{\mu} u^{\nu} \), the label \(G\) indicates the gauge used to evaluate the metric perturbation and the label \(F\) indicates that this expression is valid within the class of asymptotically flat gauges. In~\cite{Akcay:2012}, \(z_1(\Omega)\) was calculated in Lorenz gauge and the following gauge transformation can be used to link the asymptotically flat \(h^{R,\,F}_{uu} \) metric perturbation to its Lorenz-gauge counterpart \(h^{R,\,L}_{uu} \):
\begin{equation}
h^{R,\,F}_{uu} = h^{R,\,L}_{uu} + 2q\frac{x(1-2x)}{\left(1-3x)\right)^{3/2}}.
\label{flgauge}
\end{equation}
\noindent  Hence, inserting Eq.~\eqref{flgauge} into Eq.~\eqref{detinv} leads to
 \begin{equation}
z_1(\Omega)= \sqrt{1-3x}\left(1- \frac{1}{2} h^{R,\,L}_{uu}  - 2q\frac{x(1-2x)}{\left(1-3x)\right)^{3/2}} + q \frac{x}{1-3x}\right).
\label{detinvLOR}
\end{equation}
\noindent The numerical data obtained in~\cite{Akcay:2012} for \(h^{R,\,L}_{uu} \) from \(x>1/5\) was new and the numerical accuracy for \(x<1/5\) was also much improved compared to previous results~\cite{Detweiler:2008,baracknewphi}. The EoB potential \(a(x)\) can be constructed from $h_{uu}^{R,L}$ via
\begin{equation}
 a(x) = -\frac{1}{2}\left(1-3x\right)\tilde{h}^{R,\,L}_{uu} - 2x \sqrt{1-3x},
 \label{formal_a}
 \end{equation}
 \noindent with \(\tilde{h}^{R,\,L}_{uu}= q^{-1}h^{R,\,L}_{uu}\). In~\cite{Akcay:2012}, a global fit formula for \(a(x)\) was given a dense sample of numerical values over the entire range \(0 < x < \frac{1}{3}\). The  numerical fit for \(a(x)\) that we use in this study is taken from Eq.(54) of~\cite{Akcay:2012}. 
 This global analytic fit for \(a(x)\) can be recast using the relation
 \begin{equation}
 a(x)= 2x^3\, \frac{(1-2x)}{\sqrt{1 - 3 x}}\,a_{E}(x).
 \label{pot}
 \end{equation}
Using the above dictionary, the model for \(a(x)\) in this paper reproduces the numerical data points for the function \(a_{E}(x)\) to within a maximal absolute difference of \(1.2\times10^{-5}\) over the domain \(0<x<\frac{1}{3}\). The corresponding self-force corrected energy and angular momentum are given by~\cite{Akcay:2012, barus}
\begin{eqnarray}
\label{enofxeq}
E(u(x))&=&E_0(x) +  \eta\left(-\frac{1}{3}\frac{x}{\sqrt{1-3x}}a'(x) + \frac{1}{2}\frac{1-4x}{\left(1-3x\right)^{3/2}} a(x) -E_0(x)\left( \frac{1}{2}E_0(x) + \frac{x}{3}\frac{1-6x}{\left(1-3x\right)^{3/2}} \right)\right),\\
\label{lzofxeq}
L_z (u(x))&=& L_0(x) +  \eta\left( -\frac{1}{3}\frac{x}{\sqrt{x(1-3x)}}a'(x) -\frac{1}{2}\frac{1}{\sqrt{x}\left(1-3x\right)^{3/2}} a(x)    -\frac{1}{3 }\frac{1-6x}{ \sqrt{x} \left(1-3x\right)^{3/2}}\left(E_0(x)-1\right)  \right)\,,\\
\label{rofx}
&&{\rm{with}}\qquad  \qquad u(x)= x\left( 1+ \eta\Bigg[\frac{1}{6}a'(x) + \frac{2}{3}\left(\frac{1-2x}{\sqrt{1-3x}} -1 \right) \Bigg] \right)\,,
\end{eqnarray}
\noindent where \(E_0(x)\) and \(L_0(x)\) are given by
\begin{eqnarray}
E_0(x) &=&  \frac{1-2x}{\sqrt{1 - 3 x}} -1,\\
\label{enofx_0}
L_0(x)&=&   \frac{1}{\sqrt{x (1 - 3 x)}}.
\label{lzofx_0}
\end{eqnarray}
   
\begin{figure*}[ht]
\centerline{
\includegraphics[height=0.33\textwidth,  clip]{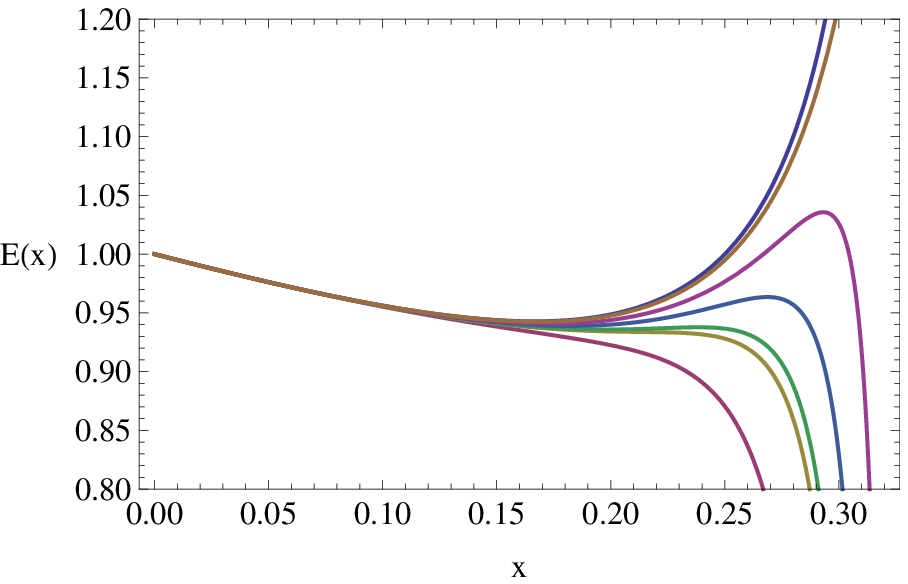}
\includegraphics[height=0.33\textwidth,  clip]{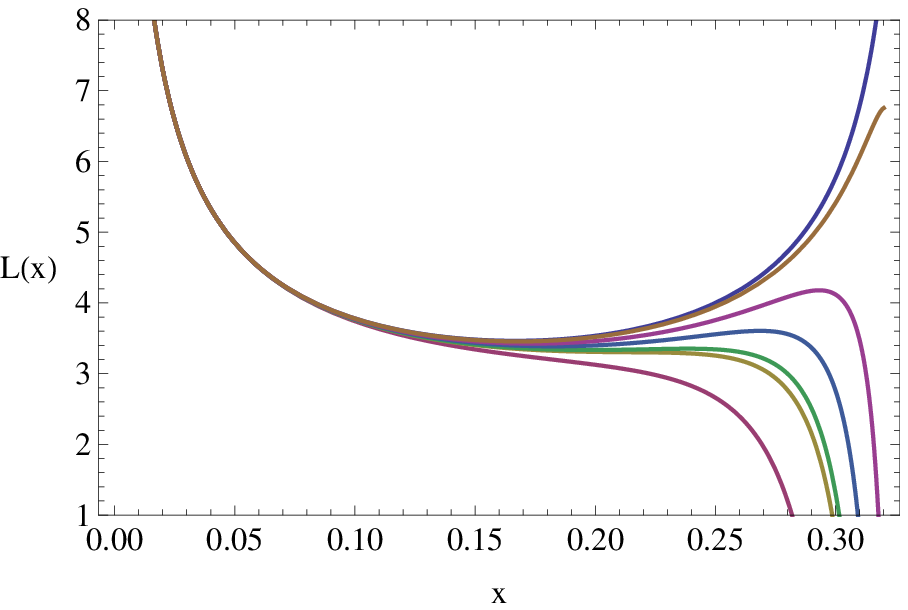}
}
\caption{The panels show the energy and angular momentum given by Eqs.~\eqref{enofxeq}-\eqref{lzofxeq}, respectively. We show the functional form of these parameters for binary systems with mass-ratio values, from top to bottom, \(q \in [0,\, 1/100, \,1/20, \,1/10, \,1/6, \,1/5, \,1 ]\).  }
\label{orbitalparams}
\end{figure*}

\noindent In Figure~\ref{orbitalparams}, we  show the effect of these conservative corrections on the orbital parameters.

\begin{figure*}[ht]
\centerline{
\includegraphics[height=0.33\textwidth,  clip]{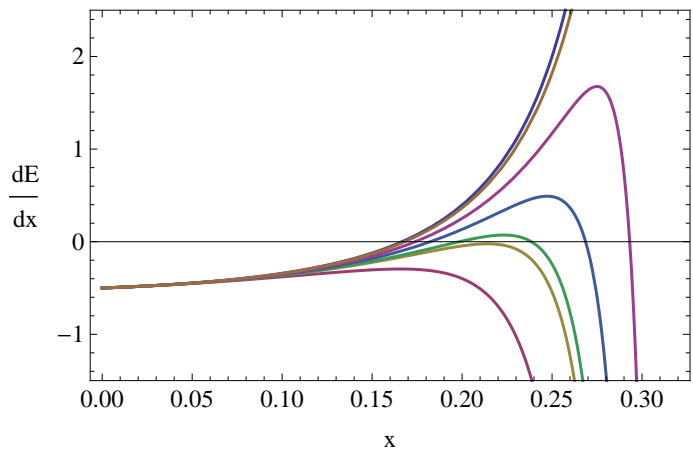}
}
\caption{The location of the innermost stable circular orbit is determined by the condition \({\rm{d}}E/\rm{d}x = 0\). The panel shows \({\rm{d}}E/\rm{d}x\) as a function of the gauge invariant quantity \(x=\left(M\,\Omega\right)^{2/3}\). The various curves represent binary systems with mass-ratios, from top to bottom, \(q \in [0,\, 1/100, \,1/20, \,1/10, \,1/6, \,1/5, \,1 ]\). Note that binaries with mass-ratios \(q\gtrsim 1/6\) do not have an ISCO in this model. }
\label{dedx}
\end{figure*}

As discussed in~\cite{barus}, if one makes use of the self-force expression for the energy given by Eq.~\eqref{enofxeq}, and minimize it with respect to the orbital frequency, then one finds that binary systems with mass-ratios \(q\in [1, \,1/2, \,1/3]\) do not have an ISCO in this model. It was argued in~\cite{barus} that deriving self-force results in the strong field regime may alleviate this problem. We have explored this issue, and have found that using linear self-force corrections that are valid all the way to the light ring (last unstable circular orbit for massless particles) does not fix this problem for comparable mass-ratio systems. In Figure~\ref{dedx}, we show that the existence of an ISCO is guaranteed for BH binaries with symmetric mass-ratio \(\eta\lesssim 6/49\, ({\rm{or}}\, q \lesssim 1/6)\), and its location may be approximated by
\begin{equation}
x_{\mathrm{ISCO}}=\frac{1}{6}\left(1+ 0.83401\eta+4.59483\eta^2\right).
\label{xisco_eq}
\end{equation}
\noindent It remains to be seen whether the inclusion of second-order conservative corrections gives an ISCO for binaries with mass-ratios \(q\gtrsim 1/6\).

In summary, the building blocks we use to construct the conservative dynamics are

\begin{itemize}
\item The orbital frequency evolution is computed using Eq.~\eqref{new_phase} with the gravitational self-force contribution included in the potential \(A(u)= 1-2u + \eta\, a(u)\).
\item Eq.~\eqref{new_phase} is evaluated using the self-force-corrected expression for the angular momentum, \(L(x)\), given by Eq.~\eqref{lzofxeq}. The self-force-corrected expression for the energy, given in Eq.~\eqref{enofxeq} is only used to determine the point at which the inspiral ends and the transition region begins.
\end{itemize}

Eq.~\eqref{new_phase} provides a very accurate modeling of the orbital frequency from early inspiral through the ISCO. However, the post-ISCO time evolution of this prescription does not render an accurate representation of the orbital frequency as compared to numerical relativity simulations. This is a problem that has been addressed in the EOB formalism by introducing a phenomenological approach ---the so-called non-quasi-circular coefficients--- that enabled them to reproduce the orbital evolution extracted from numerical simulations~\cite{buho}. The approach we will follow to circumvent this problem is described in detail in Section~\ref{trans}, and consists of embedding the self-force formalism in the context of the implicit rotating source model~\cite{Baker:2008} after the ISCO.

This completes the description of the conservative part of the self-force. We now describe how to couple this with the radiative part of the self-force to model the inspiral evolution. 
 
 \subsection{Dissipative dynamics}

A consistent self-force evolution model that incorporates first-order in mass-ratio conservative  corrections should also include second-order radiative corrections. The model we described in the previous section was constructed including first order conservative corrections. However, second-order self-force radiative corrections are not known at present. Several studies have demonstrated the importance of including the missing second order corrections to the radiative part of the self-force, both for source detection and for parameter estimation~\cite{Isoyama:2013, Burko:2012, Huerta:2012, Huerta:2010, Huerta:2009}. 

We use a new prescription for the energy flux that uses the first-order in mass ratio terms derived in~\cite{Fujita:2012}, including PN corrections up to \(22^\mathrm{nd}\) PN order
\begin{eqnarray}
\label{enfluxpn}
\left(\dot E\right)_{\rm PN} &=& -\frac{32}{5}\frac{\mu^2}{M}x^{7/2}\Bigg[1-\frac{1247}{336}x + 4\pi x^{3/2}  - \frac{44711}{9072}x^2-\frac{8191}{672}\pi x^{5/2}  \\\nonumber &+&x^3 \bigg\{\frac{6\,643\,739\,519}{69\,854\,400} +\frac{16}{3}\pi^2 -\frac{1712}{105}\gamma_{\rm E} -\frac{856}{105}\ln(16x) \Big\} -\frac{16285}{504} \pi x^{7/2} \\\nonumber&+&  x^4 \Big\{-\frac{323105549467}{3178375200}  + \frac{232597}{4410}\gamma_{\rm E} -\frac{1369}{126}\pi + \frac{39931}{294}\ln(2)  -\frac{47385}{1568}\ln(3)  +\frac{232597}{4410}\ln(x) \Big\}  \\\nonumber &+&  x^{9/2}\Big\{ \frac{265978667519}{745113600}\pi -\frac{6848}{105}\gamma_{\rm E}\pi -\frac{13696}{105}\pi\ln(2)  -\frac{6848}{105}\pi\ln(x)\Big\} \\\nonumber&+& {\rm{ higher \, order \, corrections \, up\, to \, 22PN \, order}} \Bigg].
\end{eqnarray}

\begin{figure*}[ht]
\centerline{
\includegraphics[height=0.35\textwidth,  clip]{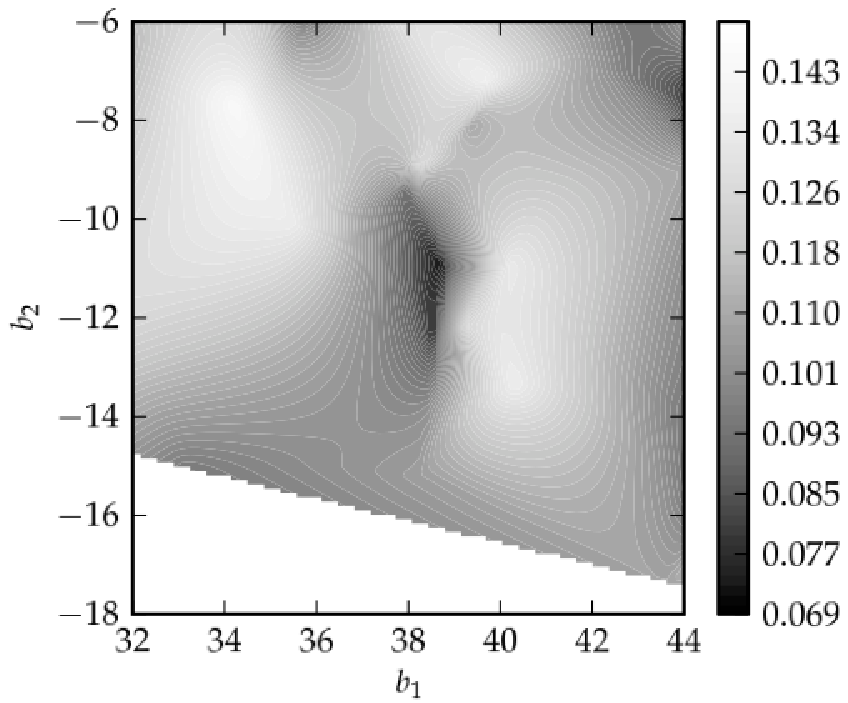}
\includegraphics[height=0.35\textwidth,  clip]{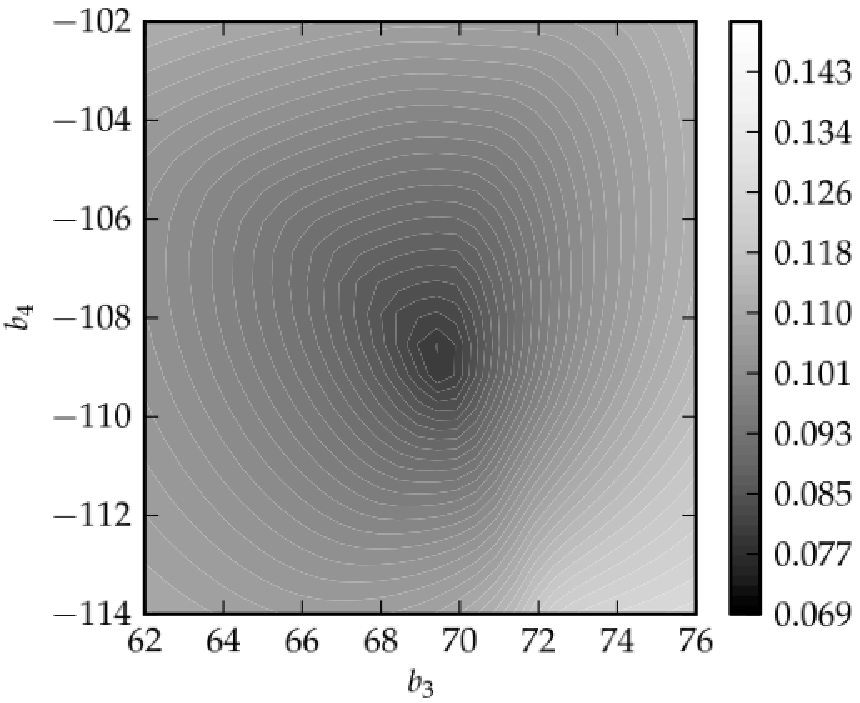}
}
\centerline{
\includegraphics[height=0.35\textwidth,  clip]{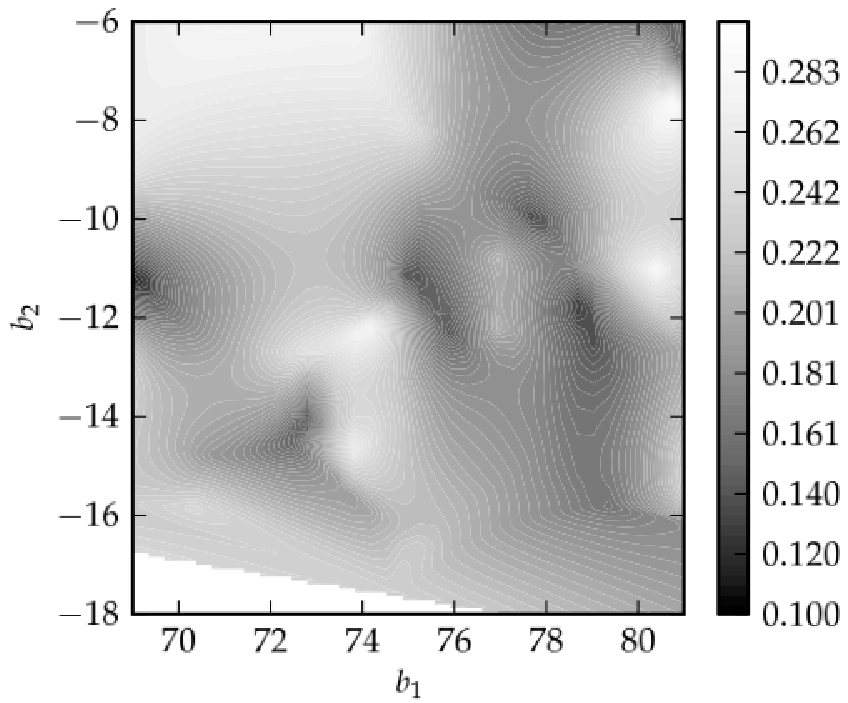}
\includegraphics[height=0.35\textwidth,  clip]{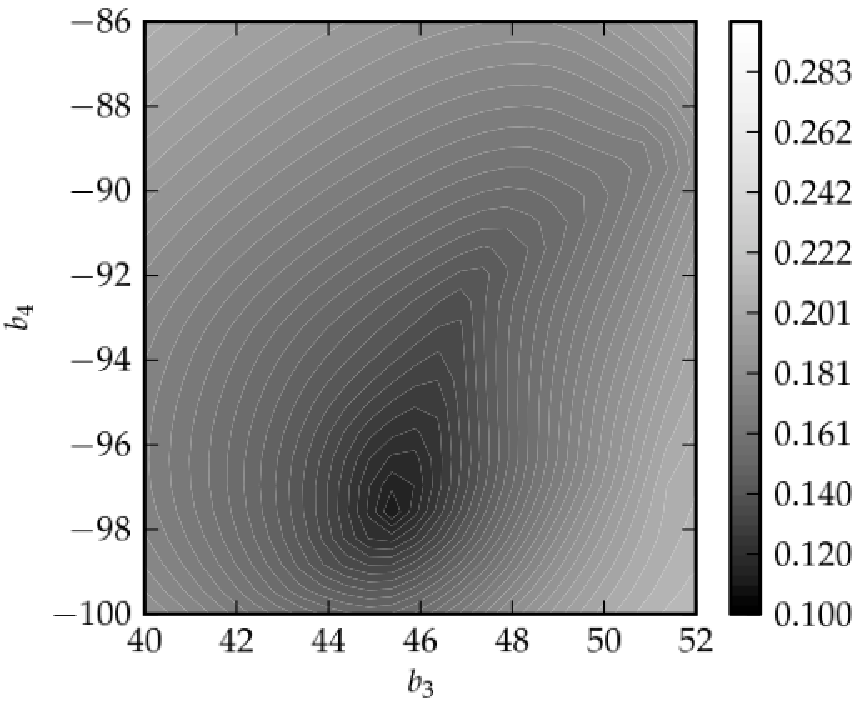}
}
\caption{The (top, \, bottom) panels show the results of the optimization runs that were used to constrain the values of the \(b_i\) coefficients given in Eq.~\eqref{etacorrect}. The panels show the results for binaries of mass--ratio \(q\in [1/6,\, 1/8]\), and total mass \(M= [7M_{\odot},\,   9M_{\odot}]\). The `optimal' value for the coefficients has been chosen by ensuring that the flux prescription minimizes the phase discrepancy between the EOB model and our self-force model. The color bar shows the phase difference squared between both models, which is integrated from \(r=20M\) all the way to the light-ring.} 
\label{bimaps}
\end{figure*}

We include higher-order in mass-ratio terms using the exponential resummation approach described in~\cite{Isoyama:2013}. In this approach, the energy flux is
\begin{eqnarray}
\left(\frac{{\mathrm{d}}E}{{\mathrm{d}}t}\right)_{\rm hybrid} &=& {\cal{L}}_{\rm 0} \exp\left( {\cal{L}}_{\eta} \right)\, ,
\label{pnfluxfit}
\end{eqnarray}
\noindent where \( {\cal{L}}_{\rm 0} \) denotes the leading-order in mass-ratio PN energy flux given in Eq.~\eqref{enfluxpn}, and \( {\cal{L}}_{\eta} \) incorporates mass-ratio corrections to the highest PN order available~\cite{Joguet:2002,Buonanno:2011_tail,Isoyama:2013}, and additional corrections characterised by a set of unknown coefficients,  \(b_i\) 
\begin{eqnarray}
\label{etacorrect}
{\cal{L}}_{\eta} &=& \Bigg[x\bigg[-\frac{35}{12}\eta + b_1\,\eta^2\bigg] + 4\pi x^{3/2}\bigg[ b_2\, \eta + b_3 \eta^2 \bigg]  + x^2 \bigg[\frac{9271}{504}\eta + \frac{65}{18}\eta^2\bigg]  + \pi x^{5/2}\bigg[ -\frac{583}{24} \eta + b_4\, \eta^2\bigg] \\\nonumber &+& x^3 \bigg[\eta\left(-\frac{134\,543}{7\,776} + \frac{41}{48}\pi^2\right) -\frac{94403}{3024}\eta^2 - \frac{775}{324}\eta^3\bigg] +  \pi x^{7/2} \bigg[\frac{214745}{1728} \eta +  \frac{193385}{3024}\eta^2\bigg]   \Bigg].
\end{eqnarray}
\noindent  The coefficients \(b_i\) were taken to be constant in~\cite{Isoyama:2013}, but we found that a better match to the EOB phase evolution could be obtained by allowing an additional dependence on mass-ratio in these terms (see Eqs.~\eqref{B1}-\eqref{B3} below).  We constrain the \(b_i\) coefficients by ensuring that the phase evolution of this model reproduces the phase evolution predicted by the EOB model introduced in~\cite{buho, Damour:2013}, which was calibrated to the phase evolution of compact binaries observed in numerical relativity simulations. To do so, we implemented the EOB model~\cite{buho} and performed a Monte Carlo simulation to optimize the values of the \(b_i\) coefficients (see Figure~\ref{bimaps}). The optimization was done in two stages. We started by considering the three coefficients \(b_1,\, b_2\) and \(b_4\),  sampling a wide range of parameter space, namely \(b_i\in[-200,200]\). We constrained the duration of the waveform from early inspiral to the light-ring to be similar to its EOB counterpart. Waveforms that differed from their EOB counterparts by more than \(10^{-4}\) seconds were discarded.  Once the region under consideration had been sparsely sampled, we focused on regions of parameter space where the orbital phase evolution was closest to the EOB evolution, and finely sampled these to obtain the optimal values for the coefficients. We found that this approach enabled us to reproduce the EOB phase evolution with a phase discrepancy of the order \(\sim 1\) rad. After constraining \(b_1,\, b_2\) and \(b_4\), we explored whether including additional corrections could further improve the phase evolution, by adding \(\eta\) corrections beyond 3PN order. Such corrections were found to have a negligible impact on the actual phase evolution. This is not difficult to understand, since such corrections are of order \(({\cal{O}}(\eta^4),\, {\cal{O}}(\eta^3))\), at (3PN, 3.5PN) respectively.  We found a similar behavior when we added leading order mass-ratios corrections beyond 4PN order. Thus, we took a different approach: having derived the optimal value for \(b_1,\, b_2\) and \(b_4\), we took these results as initial seeds for an additional MC simulation in which \(b_3\) was also included in Eq.~\eqref{etacorrect}, and repeated the optimization procedure. The results of these simulations are shown in Figure~\ref{bimaps}.

We carried out several different Monte Carlo runs to find the `optimal' optimization interval, meaning the range of radial separations over which we tried to best match the phase evolution relative to the EOB model. We found that starting the optimization at \(r=30M\) gave results that performed moderately well at early inspiral, but that underperformed at late inspiral, leading to phase discrepancies of order \(\sim 3\) rads. Starting the optimization at \(r=20M\) instead decreased the phase discrepancy with respect to the former case by a factor of 10 during early inspiral, and enabled us to reproduce the phase evolution in the EOB model for all the mass-ratios considered to within the accuracy of the numerical waveforms used to calibrate the EOB model in~\cite{buho, Damour:2013}. Implementing these numerically optimized higher-order \(\eta\) corrections in Eq.~\eqref{etacorrect} leads to:

\begin{eqnarray}
\label{etacorrect_new}
{\cal{L}}_{\eta} &=& \Bigg[x\bigg[-\frac{35}{12}\eta + B_1\bigg] + 4\pi x^{3/2}B_2  + x^2 \bigg[\frac{9271}{504}\eta + \frac{65}{18}\eta^2\bigg]  + \pi x^{5/2}\bigg[ -\frac{583}{24} \eta + B_3\bigg] \\\nonumber &+& x^3 \bigg[\eta\left(-\frac{134\,543}{7\,776} + \frac{41}{48}\pi^2\right) -\frac{94403}{3024}\eta^2 - \frac{775}{324}\eta^3\bigg] +  \pi x^{7/2} \bigg[\frac{214745}{1728} \eta +  \frac{193385}{3024}\eta^2\bigg]   \Bigg].
\end{eqnarray}
\noindent where:

\begin{eqnarray}
\label{B1}
B_1&=& \frac{1583.650 - 11760.507\, \eta}{1 + 142.389\, \eta - 981.723\, \eta^2}\,\eta^2\,,\\
\label{B2}
B_2 &=& \frac{-12.081 + 35.482\, \eta}{1 - 4.678 \eta + 13.280\, \eta^2}\,\eta +  \frac{19.045 - 240.031\, \eta}{1 - 18.461\, \eta + 74.142\, \eta^2}\,\eta^2\,,\\
\label{B3}
 B_3 &=& \frac{51.814 - 980.100\, \eta}{1 - 13.912\, \eta + 88.797\, \eta^2}\,\eta^2\,.
 \label{new_coef}
 \end{eqnarray}

\begin{figure*}[ht]
\centerline{
\includegraphics[height=0.35\textwidth,  clip]{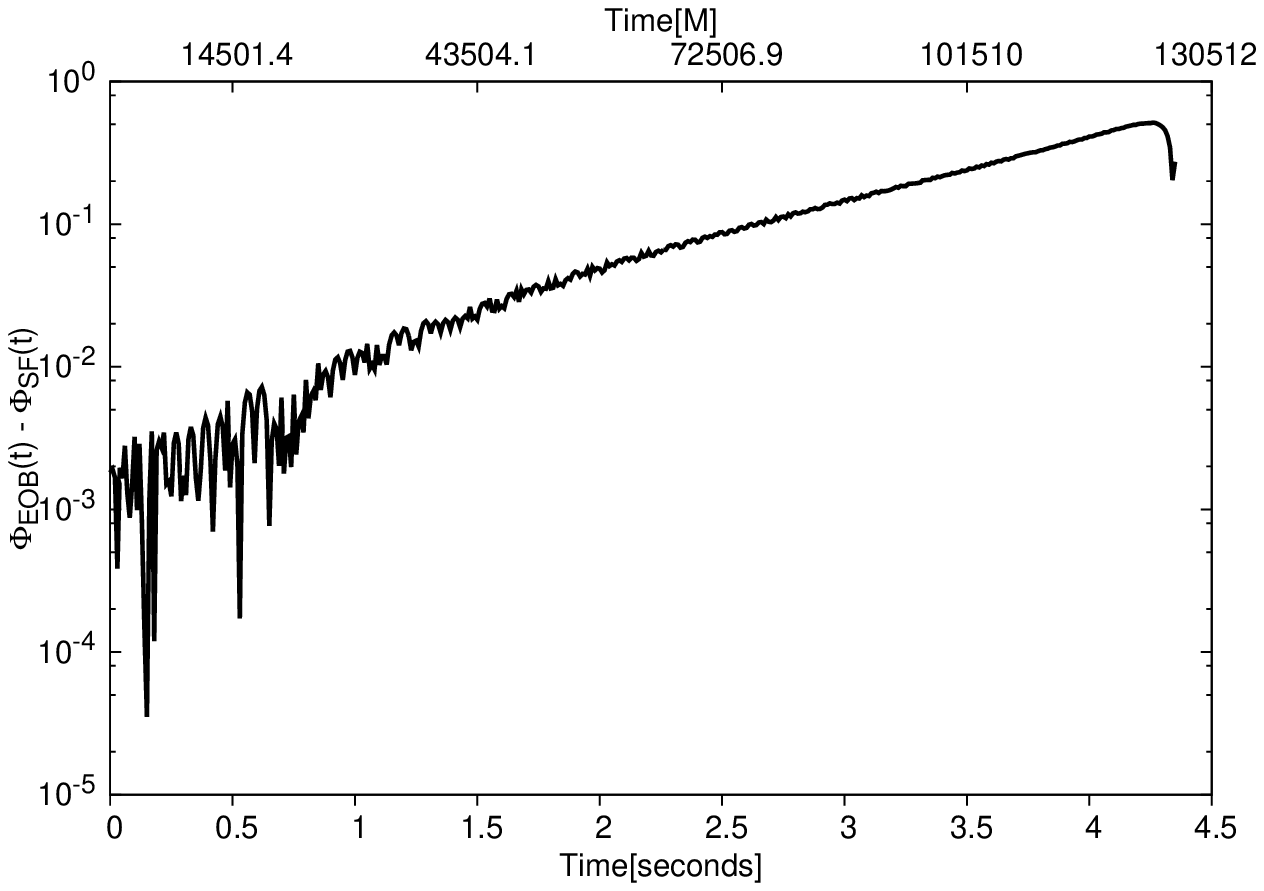}
\includegraphics[height=0.35\textwidth,  clip]{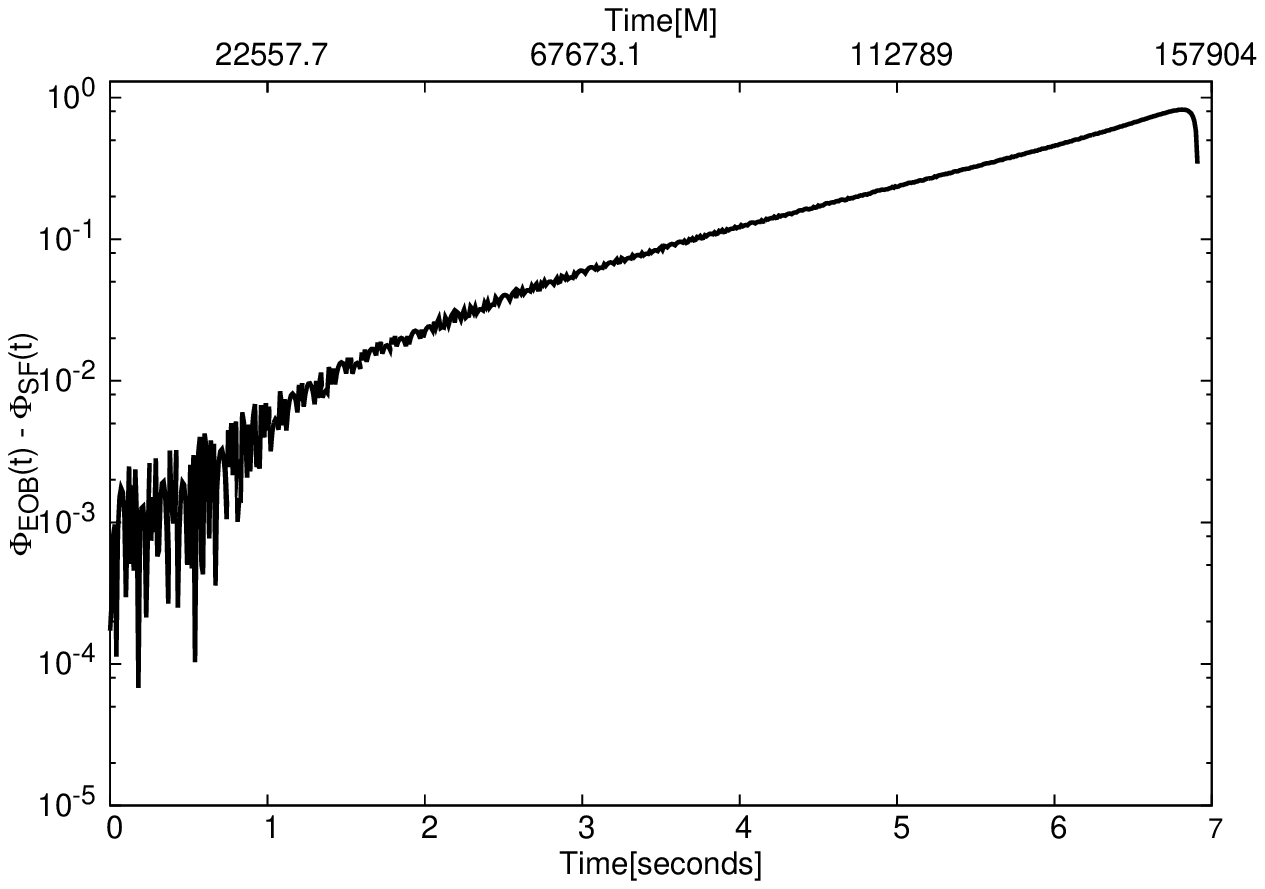}
}
\centerline{
\includegraphics[height=0.35\textwidth,  clip]{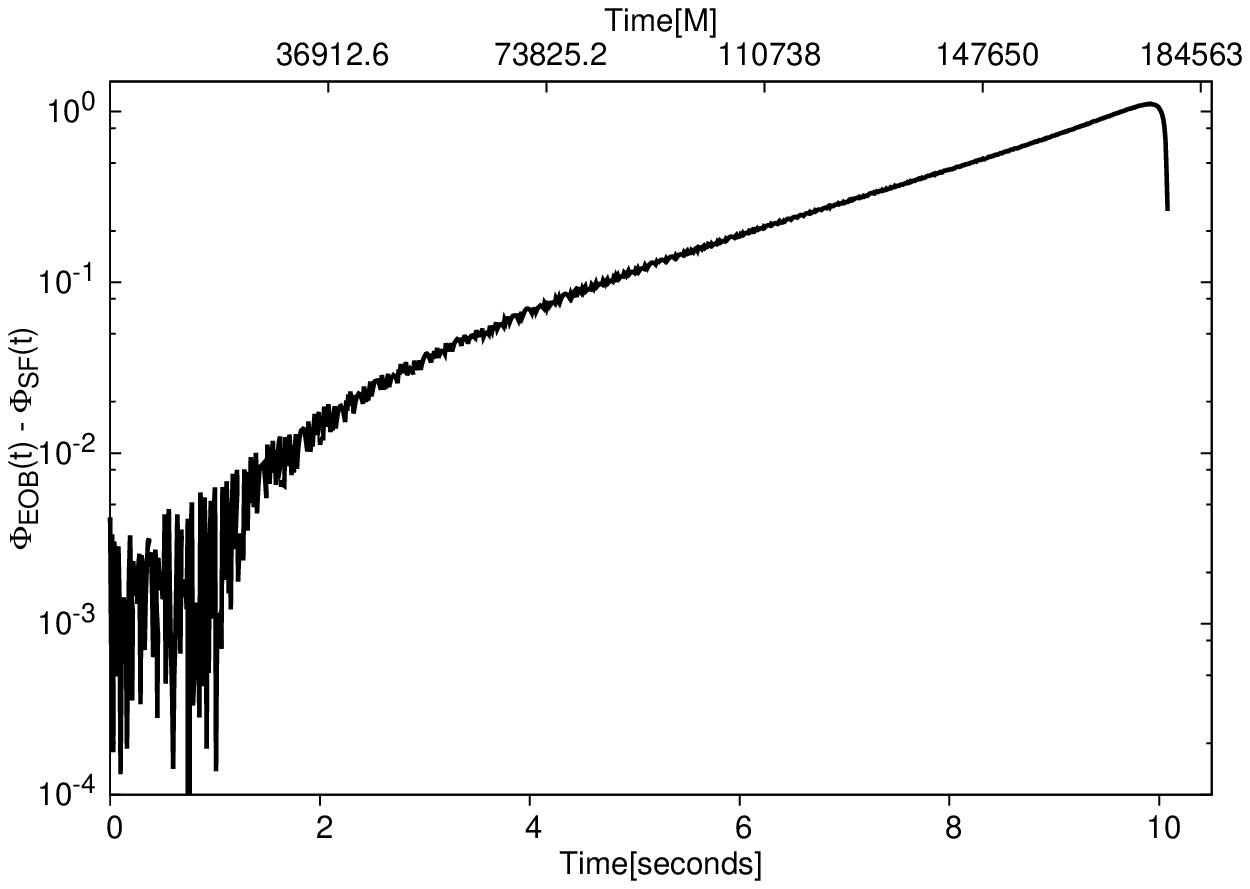}
\includegraphics[height=0.35\textwidth,  clip]{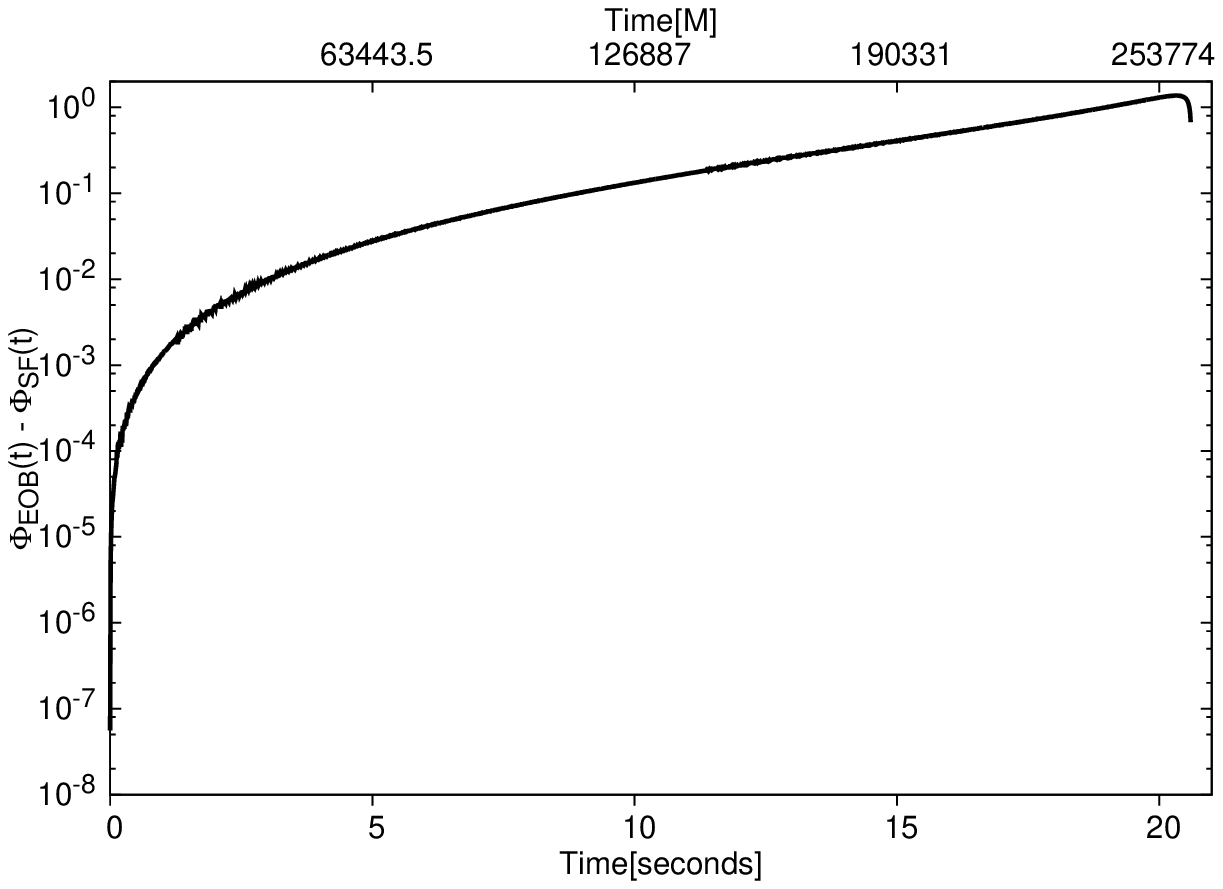}
}
\caption{The panels show the orbital phase evolution of a self-force model making use of optimized PN energy flux given by Eq.~\eqref{pnfluxfit} and the phase evolution as predicted by the EOB model. The [top/bottom] panels exhibit this evolution for a compact binary with mass--ratio \(q=[(1/6,\,1/8), \,( 1/10,\,1/15)]\), and total mass \(M=[ (7M_{\odot} ,\, 9M_{\odot} ), \, ( 11M_{\odot},\, 16M_{\odot})  ]\), respectively. }
\label{PNoptimized}
\end{figure*}

This improved prescription for the energy flux, which incorporates second-order mass-ratio corrections to the PN expansion up to 3.5PN order, is sufficient to generate a model whose phase evolution reproduces with excellent accuracy the phase evolution predicted by EOB throughout inspiral and merger (see Figure~\ref{PNoptimized}).

Given the energy flux defined by Eqs~\eqref{enfluxpn}--\eqref{etacorrect}, we generate the inspiral trajectory using the simple prescription 
\begin{equation}
\frac{{\mathrm{d}}x}{{\mathrm{d}}t}= \frac{{\mathrm{d}} E}{{\mathrm{d}} t}\frac{{\mathrm{d}} x}{{\mathrm{d}}E}\,,
\label{radev}
\end{equation}

\noindent where we have used the mass-ratio corrected energy ---Eq.~\eqref{enofxeq}--- to compute \({\mathrm{d}}E/{\mathrm{d}}x\). Figure~\ref{PNoptimized} shows that for binaries with mass-ratio \(q=1/6\), the phase discrepancy between our self-force model and EOB is \(\lesssim 0.5\) rads at the light-ring, which is within the numerical accuracy of the simulations used to calibrate EOB. It has been shown recently that EOB remains accurate for mass-ratios up to \(q=1/8\)~\cite{Pan:2013}. In that regime the phase discrepancy between this model and EOB at the light-ring is less than 1 rad, as shown in Figure~\ref{PNoptimized}. For binaries with \(q=1/10\), the phase discrepancy at the light-ring is \(\lesssim 1.2\) rads, which is still within the numerical accuracy of available simulations~\cite{carlosI, carlosII}. 

It must be emphasized that even if we only use the inspiral evolution to model binaries with mass-ratios that typically describe NSBH binaries, our self-force evolution model performs better than TaylorT4, since we can reduce the phase discrepancy between TaylorT4 and EOB at the last stable circular orbit by a factor of \((\sim40, \, \sim70)\)  for binaries with \(q=(1/6,\,1/10)\)  and total mass \(M\in (7M_{\odot} ,\, 11M_{\odot} )\) (see Figure~\ref{pn_approx}).

Figure~\ref{PNoptimized} conveys an important message --- deriving the second order corrections to the radiative part of the self-force may well provide a robust framework to describe in a single model not only events that are naturally described by BHPT, such as the mergers of stellar mass compact objects with supermassive BHs in galactic nuclei~\cite{Huerta:2012, Huerta:2010, wargar, cutler, gairles, SFB, GairL:2013}, but also events that are more naturally described by PN or numerical methods, in particular the coalescences of compact object binaries with comparable or intermediate mass-ratios~\cite{Huerta:2012, higherspin, Huerta:2011a, Huerta:2011b, smallbody}.

To finish this Section, we describe the approach followed to construct the gravitational waveform from the inspiral trajectory. At leading PN order, a general inspiral waveform can be written as
\begin{equation}
h(t) = -(h_{+} - i h_{\times}) = \sum_{\ell=2}^{\infty} \sum_{m=-\ell}^{l} h^{\ell m} {}_{-\!2}Y_{\ell m}(\iota,\Phi).
\label{inspwav}
\end{equation}

\noindent If only the leading-order modes \((\ell,m)=(2, \pm 2)\) and included, the inspiral waveform components  are given by 
\begin{eqnarray}
h_{+}(t)&=& \frac{4\, \mu\,  r^2\, \dot{\phi}^2 }{D}\left(\frac{1+\cos^2 \iota}{2}\right)\cos\left[2(\phi(t) + \Phi)\right],\label{inspp}\\
h_{\times}(t)&=& \frac{4\, \mu\, r^2\, \dot{\phi}^2}{D} \cos\iota \sin\left[2(\phi(t) + \Phi)\right],
\label{inspc}
\end{eqnarray}
\noindent where \(D\) is the distance to the source. Since the orbital evolution will deviate from a circular trajectory during late inspiral (\(\dot{r}\neq0\)), we must consider more general orbits in which both \(\dot{r}\) and \(\dot{r}\dot{\phi}\) are non-negligible. For such orbits, the Newtonian GW polarizations are given by~\cite{Gopakumar:2002}:
\begin{eqnarray}
\label{insppcor}
h_{+}(t)&=& \frac{2 \mu }{D}\Bigg\{ \left(1+\cos^2 \iota\right) \Bigg[ \cos\left[2(\phi(t) + \Phi)\right]\left(-\dot{r}^2 + r^2 \dot{\phi}^2 + \frac{1}{r}\right) \nonumber \\ &+& 2r\,\dot{r}\,\dot{\phi}\,\sin\left[2(\phi(t) + \Phi)\right]\Bigg] + \left(-\dot{r}^2 - r^2\dot{\phi}^2 + \frac{1}{r}\right)\sin^2 \iota\Bigg\}\,,\\
\label{inspccorrected}
h_{\times}(t)&=&\frac{4 \mu }{D}\cos\iota\Bigg\{  \sin\left[2(\phi(t) + \Phi)\right]\left(-\dot{r}^2 + r^2 \dot{\phi}^2 + \frac{1}{r}\right) \nonumber \\ &-& 2r\,\dot{r}\,\dot{\phi}\,\cos\left[2(\phi(t) + \Phi)\right]\Bigg\},
\end{eqnarray}
\noindent where \(\dot{r}\) can be computed using
\begin{displaymath}
\frac{{\mathrm{d}}r}{{\mathrm{d}}t} = -\frac{1}{u^2}\frac{{\mathrm{d}u}}{{\mathrm{d}}x}\frac{{\mathrm{d}x}}{{\mathrm{d}}t}\, .
\end{displaymath}
Having described the construction of the inspiral evolution, we shall now describe the approach followed to smoothly connect the late inspiral evolution onto the plunge phase. The adiabatic prescription given by Eq.~\eqref{radev} breaks down when \(dE/dx\rightarrow 0\). Hence, we need a scheme that enables us to match the late inspiral phase onto the plunge phase. We will do this by modifying the ``transition'' phase developed by Ori and Thorne~\cite{amos} by including finite mass--ratio corrections. 

\subsection{Transition and plunge phases}
\label{trans}
\begin{figure*}[ht]
\centerline{
\includegraphics[height=0.32\textwidth,  clip]{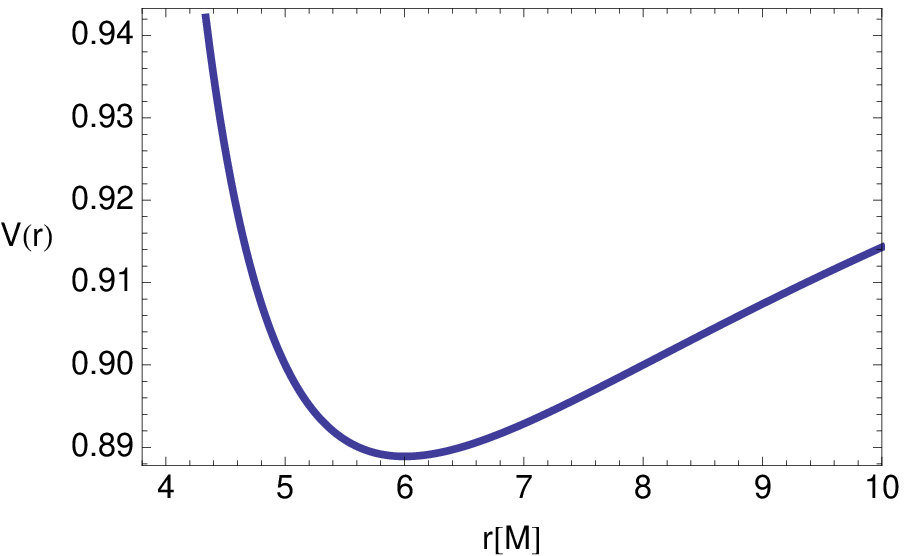}
\includegraphics[height=0.32\textwidth,  clip]{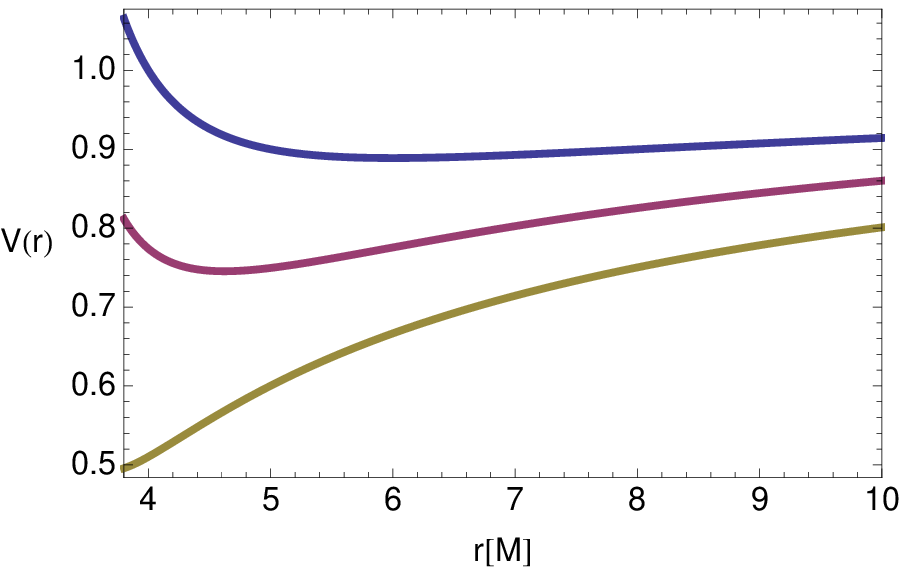}
}
\caption{Left panel: The object sits at the minimum of the effective potential, Eq.~\eqref{effpot_fmrc}, which corresponds to the case  \(\xi = L-L_{\rm{ISCO}} \gg 0\). Right panel. Blue (top) curve: radial geodesic motion, which corresponds to \(\xi = L-L_{\rm{ISCO}} \gg 0\);  Red (middle) curve: the object nears the ISCO and the orbit shrinks due to radiation reaction. Note that the minimum of the potential has moved inwards (\(\xi = 0.35\)). Yellow (bottom) curve:  body's inertia prevents it from staying at the minimum of the potential, and adiabatic inspiral breaks down  (\(\xi = 0\)). At this point the transition regime takes over the late inspiral evolution~\cite{amos}. Note: this plot is based on Figure 1 of~\cite{amos}.}
\label{effpotper}
\end{figure*}

In this Section we describe an extension of the transition phase model introduced by Ori and Thorne~\cite{amos}. The basic idea behind this approach can be understood by studying the motion of an inspiralling object in terms of the effective potential, \(V(r, L)\), which takes the following simple form for  a Schwarzschild BH~\cite{MTW}:
\begin{equation}
V(r, L) = \left(1-\frac{2}{r} \right)\left(1+ \frac{L^2}{r^2}\right).
\label{effpot_fmrc}
\end{equation} 
\noindent Throughout the inspiral, the body moves along a nearly circular orbit, and hence the radio of the energy flux to the angular momentum flux is given by:
\begin{equation}
 \frac{{\mathrm{d}}E}{{\mathrm{d}}\tau} = \Omega  \frac{{\mathrm{d}}L}{{\mathrm{d}}\tau}.
 \label{radII}
 \end{equation}
\noindent Hence, near the ISCO, the energy and angular momentum of the body satisfy the following relations:
\begin{eqnarray}
\label{ener_emri}
E & \rightarrow & E_{\rm{ISCO}} + \Omega_{\rm{ISCO}}\, \xi,\\
\label{ang_emri}
L  & \rightarrow & L_{\rm{ISCO}}+  \xi.
\end{eqnarray}
\noindent Re-writing the effective potential, Eq.~\eqref{effpot_fmrc}, in terms of \(\xi = L-L_{\rm{ISCO}}\), one notices that during early inspiral, \(\xi \gg0\), the motion of the object is adiabatic, and the object sits at the minimum of the potential ---as shown in the left panel of Figure~\ref{effpotper}. However, as the object nears the ISCO, the minimum of the potential moves inward due to radiation reaction. At some point, the body's inertia prevents the body from staying at the minimum of the potential, and adiabatic inspiral breaks down~\cite{amos} --- illustrated in the right-hand panel of Figure~\ref{effpotper}.

\noindent The equation that governs the radial motion during the transition regime is found by linearising the equation 
\begin{equation}
\left(\frac{{\mathrm{d}}r}{{\mathrm{d}}\tau}\right)^2 = E(r)^2 - V(r,\,L),
\label{geoeom}
\end{equation}
\noindent using Eqs.~\eqref{ener_emri}, \eqref{ang_emri}, and
\begin{equation}
\label{xi_eq}
\frac{{\mathrm{d}}\xi}{{\mathrm{d}}\tau}= \kappa\, \eta, \quad {\rm{with}} \quad \kappa =\bigg[\frac{32}{5}\Omega^{7/3} \frac{\dot{{\cal{E}}}}{\sqrt{1-3u}}\bigg]_{\rm{ISCO}} \,,
\end{equation}
\noindent where \(\dot{{\cal{E}}}\) is the general relativistic correction to the Newtonian, quadrupole-moment formula~\cite{ori}. We now extend these Eqs. by including finite mass-ratio corrections. Eq.~\eqref{geoeom} can be replaced by 
\begin{equation}
\frac{{\mathrm{d}}x}{{\mathrm{d}}t}= \frac{u^2(1-2u)}{E(x) }\left(\frac{{\mathrm{d}}u}{{\mathrm{d}}x}\right)^{-1} \bigg[E(x)^2 - V\left(u(x),L(x)\right)\bigg]^{1/2},
\label{eomfmrc}
\end{equation}
 \noindent where we have used
 \begin{equation}
 \frac{{\mathrm{d}}\tau}{{\mathrm{d}}t}= \frac{1-2\,u(x)}{E(x)}\,,
 \label{tfact}
 \end{equation}
\noindent and the expressions for the energy and angular momentum are given by Eqs.~\eqref{enofxeq}, \eqref{lzofxeq}. In order to linearize Eq.~\eqref{eomfmrc} we replace $E(x)$ and $L(x)$ by Eqs~\eqref{ener_emri} and \eqref{ang_emri} respectively.

\begin{figure*}[ht]
\centerline{
\includegraphics[height=0.33\textwidth,  clip]{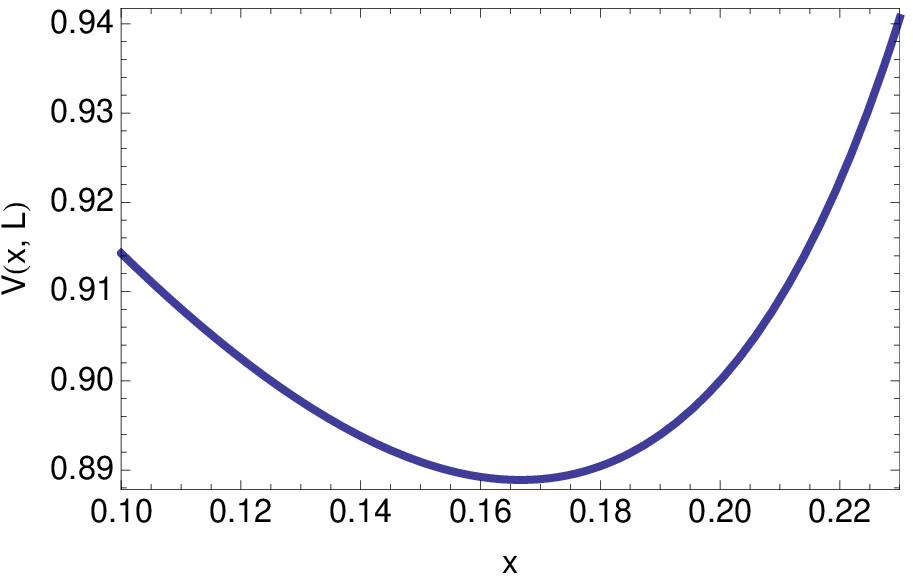}
\includegraphics[height=0.33\textwidth,  clip]{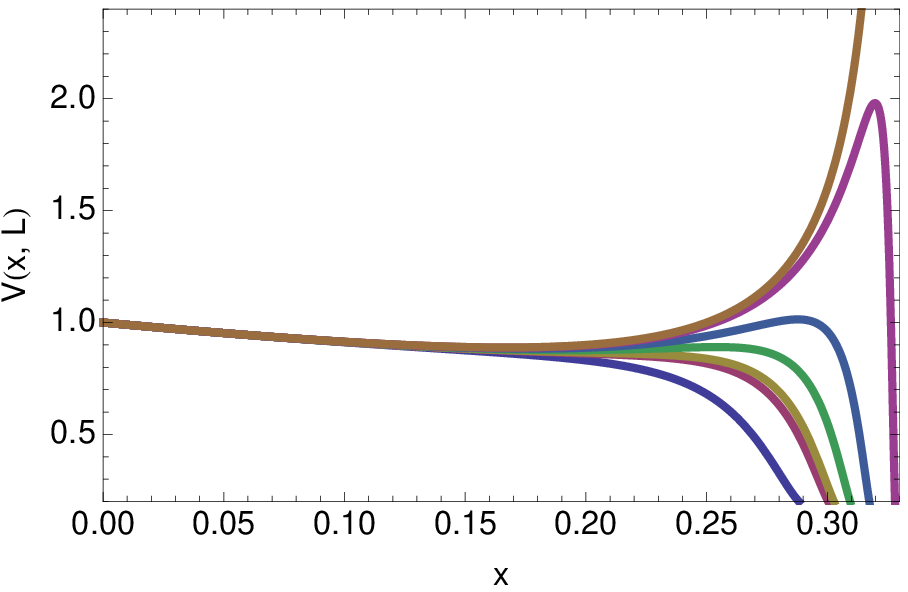}
}
\caption{The left panel shows the effective potential for a Schwarzschild BH without including finite mass--ratio corrections. Note that the minimum of the potential takes place at the ISCO, which can be determined using Eq.~\eqref{xisco_eq}.  The right panel exhibits the influence of finite mass-ratio corrections on the effective potential used to modify Ori and Thorne transition regime~\cite{amos}. The curves represent  binaries, from top to bottom, with mass-ratios  \(q \in [0,\, 1/100, \,1/20, \,1/10, \,1/6, \,1/5, \,1 ]\).}
\label{eff_pot_fig}
\end{figure*}

As discussed in~\cite{amos}, since these equations use the \(\eta\)-corrected values for \(E(x_{\rm{ISCO}} )\), \(L(x_{\rm{ISCO}}) \) and \( \Omega_{\rm{ISCO}}\), then they remain valid even for finite mass-ratio \(\eta\)~\cite{amos}. In Figure~\ref{eff_pot_fig} we show the effect that these finite mass-ratio \(\eta\) corrections have on the effective potential \(V(x, L(x))\). We determine the point at which the transition regime starts by carrying out a stability analysis near the ISCO using \({\mathrm{d}} E/ {\mathrm{d}} x\). As shown in Figure~\ref{dedx}, the ISCO is determined by the relation \({\mathrm{d}} E/{\mathrm{d}} x =0\). We have found that the relation 
\begin{equation}
\left(\frac{{\mathrm{d}}E}{{\mathrm{d}}x}\right)\Bigg |_{\rm{transition}} = -0.054 + \frac{1.757\times10^{-4}}{\eta}\,,
\label{transition_point}
\end{equation}
\noindent provides a robust criterion to mark the start of the transition regime for binaries with mass-ratios \(1/100<q<1/6\).

In~\cite{ori}, the authors only kept terms linear in \(\xi\), but we have explored which higher order terms had a noticeable impact on the evolution by examining their impact on the length  and phasing of the waveform. We found that terms  \(\propto \xi\) and \( \propto (u- u_{\rm{ISCO}})\xi\) were important, but corrections at order \({\cal{O}}(\xi^2)\) could be ignored even for comparable mass-ratio systems.

We model the evolution of the orbital frequency during the transition regime and thereafter in a  different manner to that proposed by Ori and Thorne~\cite{amos}. In order to ensure that the late-time evolution of the orbital frequency of our self-force model is as close as possible to the orbital evolution extracted from numerical relativity simulations, we incorporate the late-time frequency evolution that was derived by Baker et al~\cite{Baker:2008} in their implicit rotating source (IRS) model, namely:
\begin{equation}
\frac{d \phi}{d \mathrm{t} }  = \Omega_{\rm{i}}+ \left(\Omega_{\rm{f}}\  -\Omega_{\rm{i}}\right)\left(\frac{1 + \tanh( \ln\sqrt\varkappa + (t-t_0)/b)}{2}\right)^{\varkappa}\, ,
\label{late_frequency}
\end{equation}
\noindent where \( \Omega_{\rm{i}}\) is the value of the orbital frequency when the transition regime begins, and  \( \Omega_{\rm{f}}\) is the value of the frequency at the light ring, which corresponds to \(\omega_{\rm{\ell m n}}/m\), where \(\omega_{\rm{\ell m n}}\) is the fundamental quasi-normal ringing frequency \((n=0)\) for the fundamental mode \((\ell, m) = (2,2)\) of the post-merger black hole (see Eq.~\eqref{finspin} below).  The constant mass-dependent coefficient \(t_0\) is computed by ensuring that \(d\Omega/dt\) peaks at a time \(t=t_0\). The parameter \(\varkappa\) is computed by enforcing continuity between the first order time derivative of the orbital frequency as predicted by the self-force evolution ---Eq.~\eqref{new_phase}--- and that given by the first order time derivative of Eq.~\eqref{late_frequency}.

At the end of the plunge phase, we match the plunge waveforms, which are generated using Eqs.~\eqref{insppcor}, \eqref{inspccorrected}, onto the \(l=m=2\), \(n=0\) ringdown mode since this dominates the ringdown radiation. In the following Section we will describe in detail the procedure followed to attach the ringdowm waveform. 

After the transition regime, the equations of motion we use to model the plunge phase are:  the second order time derivative of Eq.~\eqref{eomfmrc} which gives the radial evolution, and Eq.~\eqref{late_frequency} which describes the orbital frequency evolution. We determine the point at which to attach the plunge phase by integrating  Eq.~\eqref{eomfmrc} backwards in time, and finding the point at which the transition and plunge equations of motion smoothly match.

\begin{figure*}[ht]
\centerline{
\includegraphics[height=0.36\textwidth,  clip]{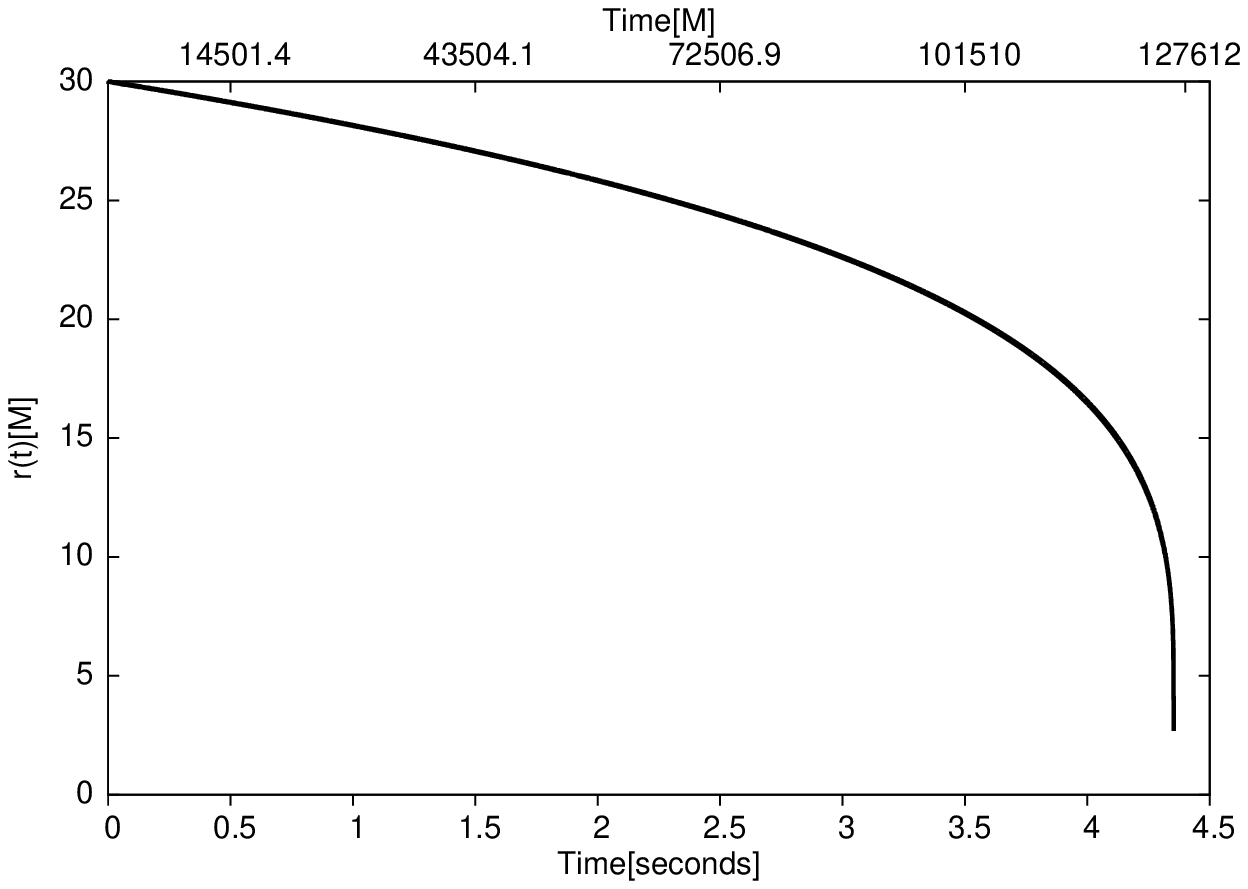}
\includegraphics[height=0.36\textwidth,  clip]{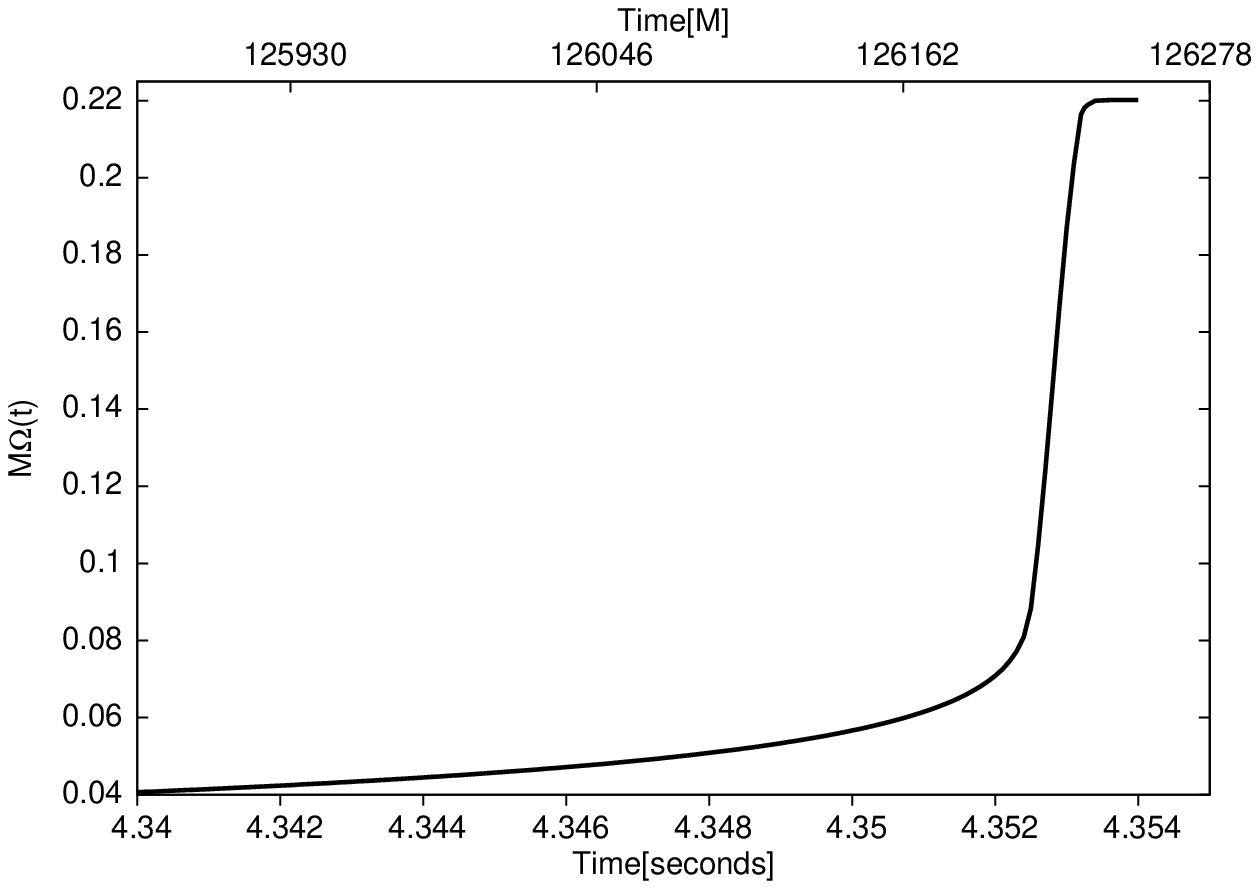}
}
\centerline{
\includegraphics[height=0.36\textwidth,  clip]{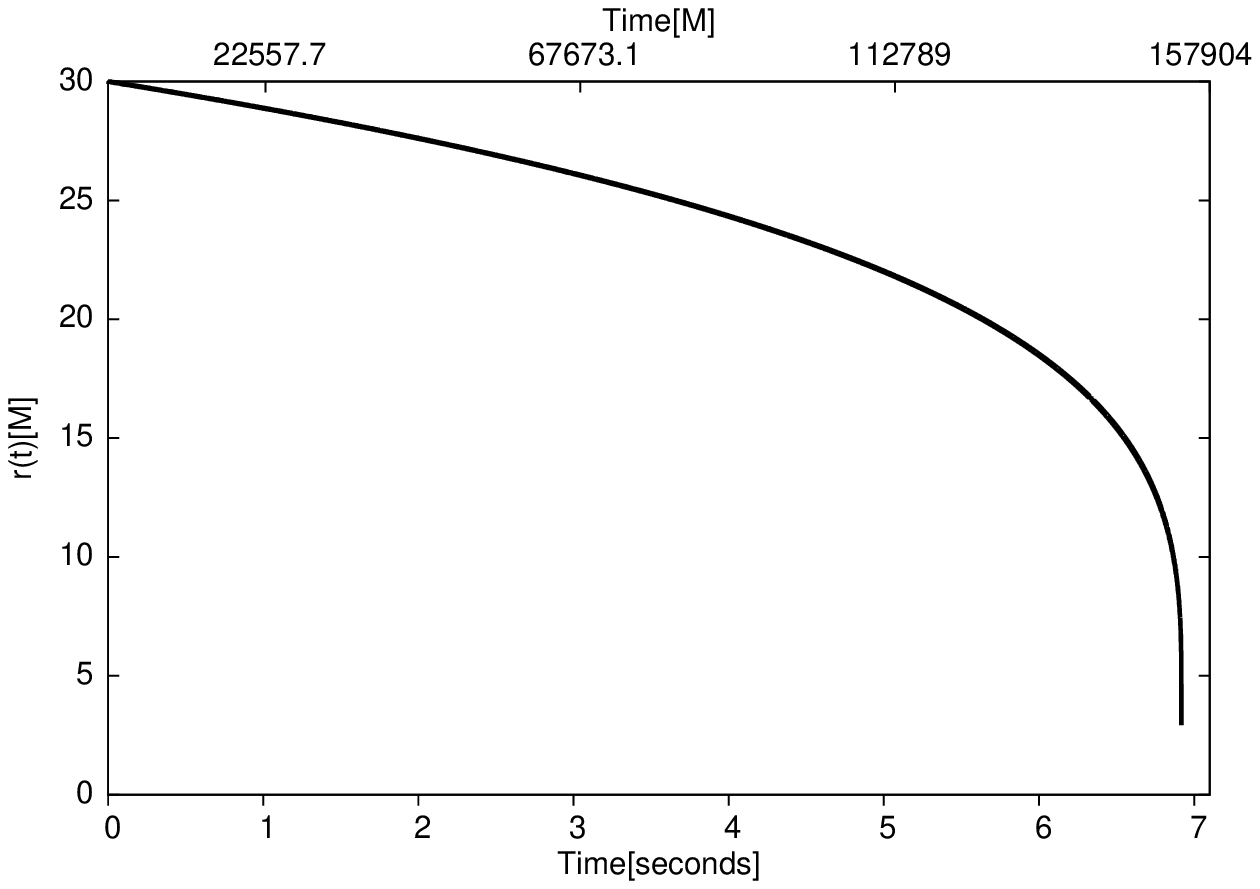}
\includegraphics[height=0.36\textwidth,  clip]{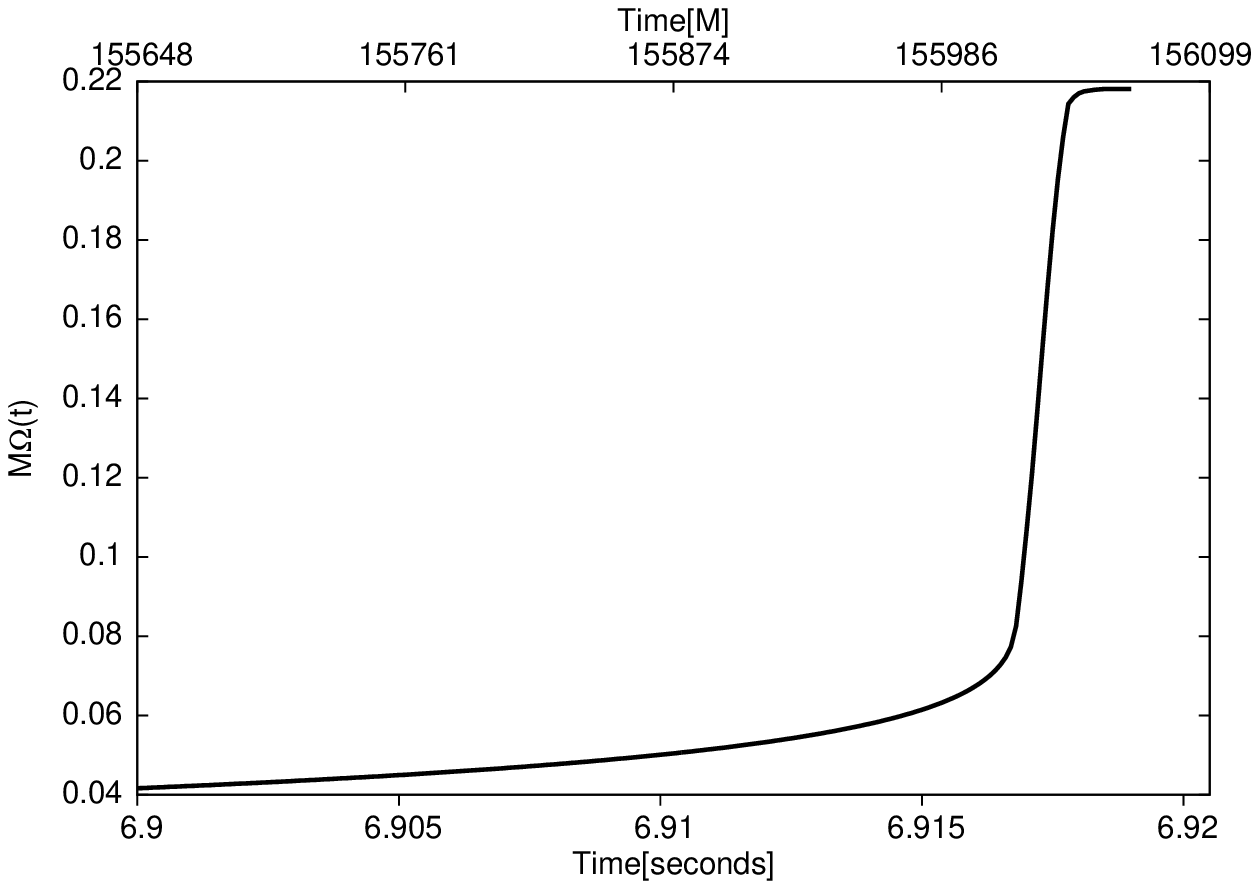}
}
\caption{(Top, bottom) panels: the left panel shows the inspiral, transition and plunge radial evolution for a BH binary of mass-ratio \(q=(1/6,\,1/8)\) --- and total mass \(M\in (7M_{\odot} ,\, 9M_{\odot} )\) --- using the coordinate transformation given by Eq.~\eqref{rofx}. The right panel shows the orbital frequency \(M\Omega\) from late inspiral all the way to the light ring. The evolution starts from an initial radial value \(r=30M\).}
\label{IP}
\end{figure*}

\begin{figure*}[ht]
\centerline{
\includegraphics[height=0.36\textwidth,  clip]{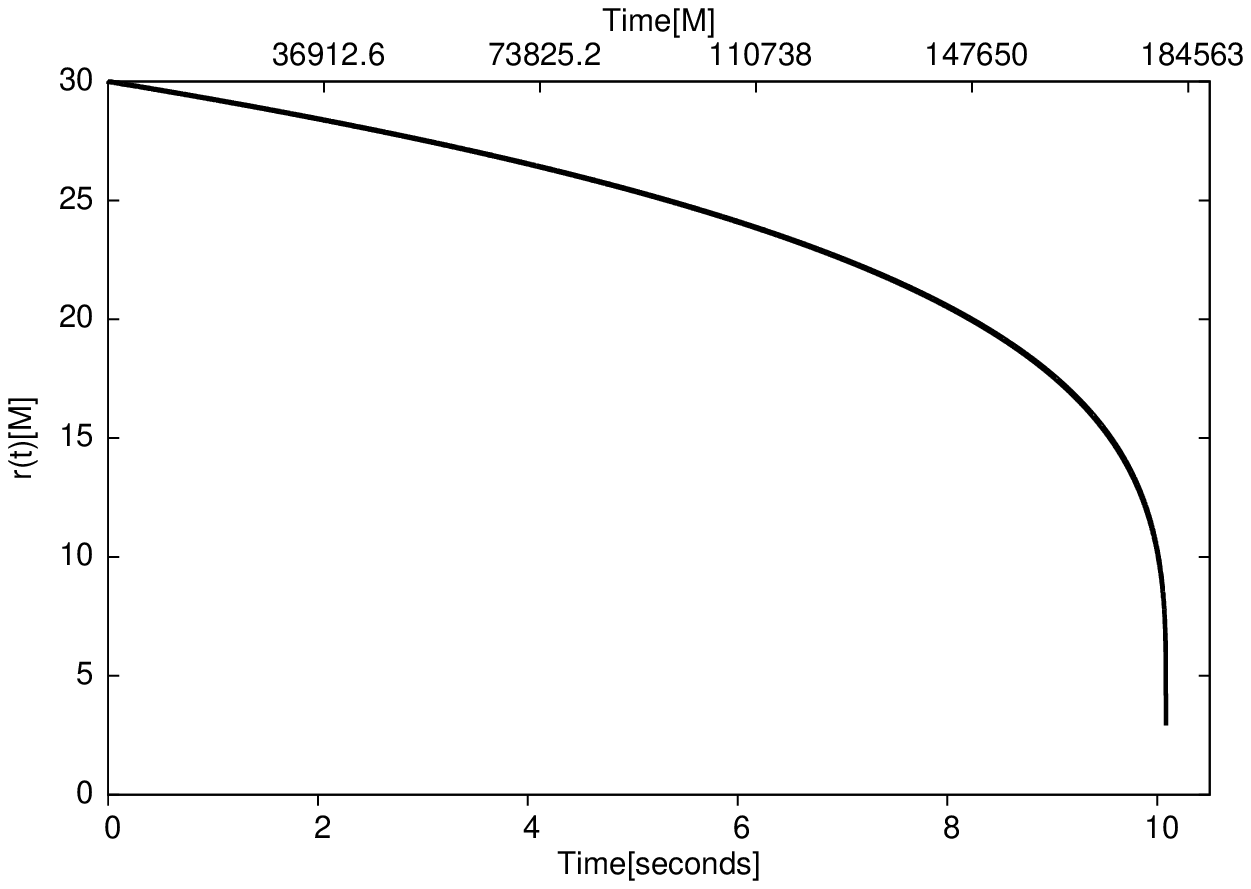}
\includegraphics[height=0.36\textwidth,  clip]{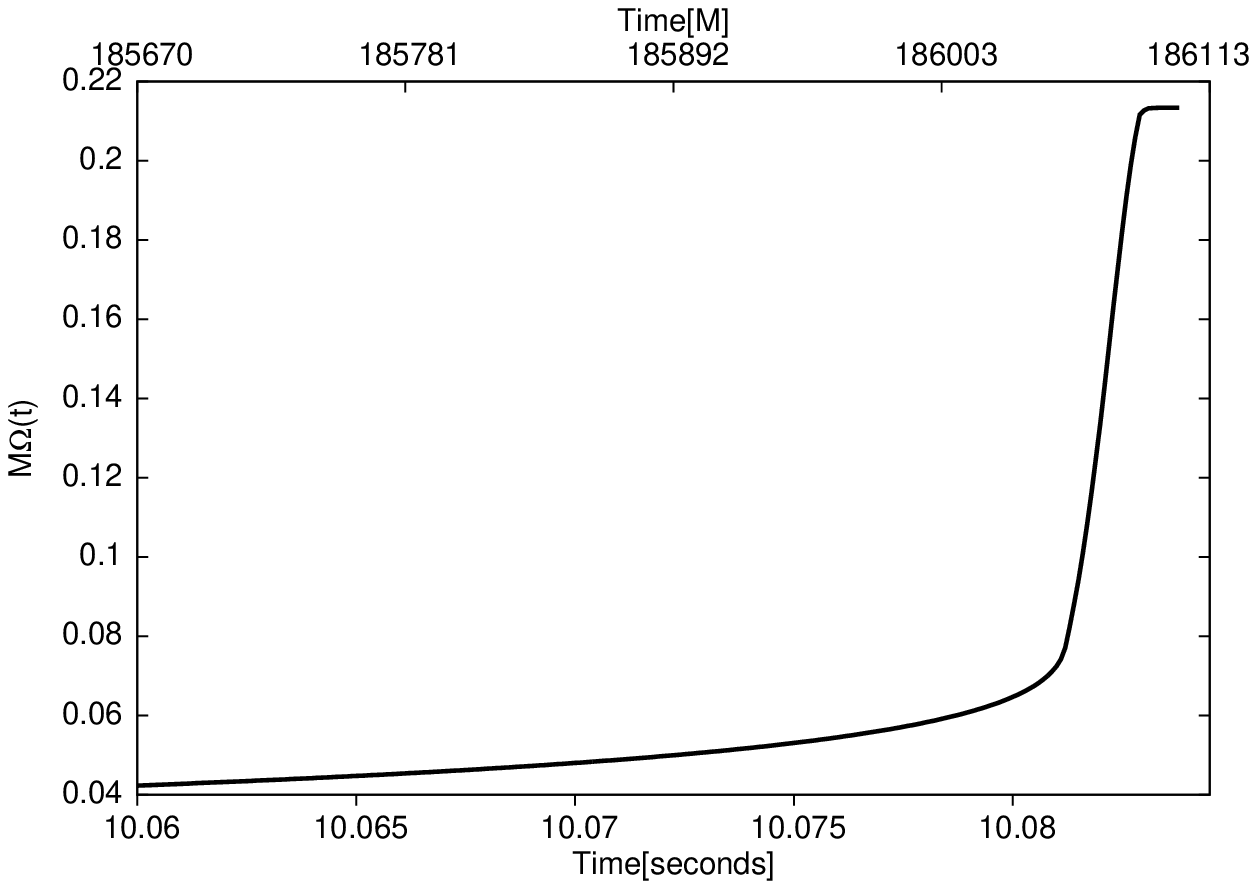}
}
\centerline{
\includegraphics[height=0.36\textwidth,  clip]{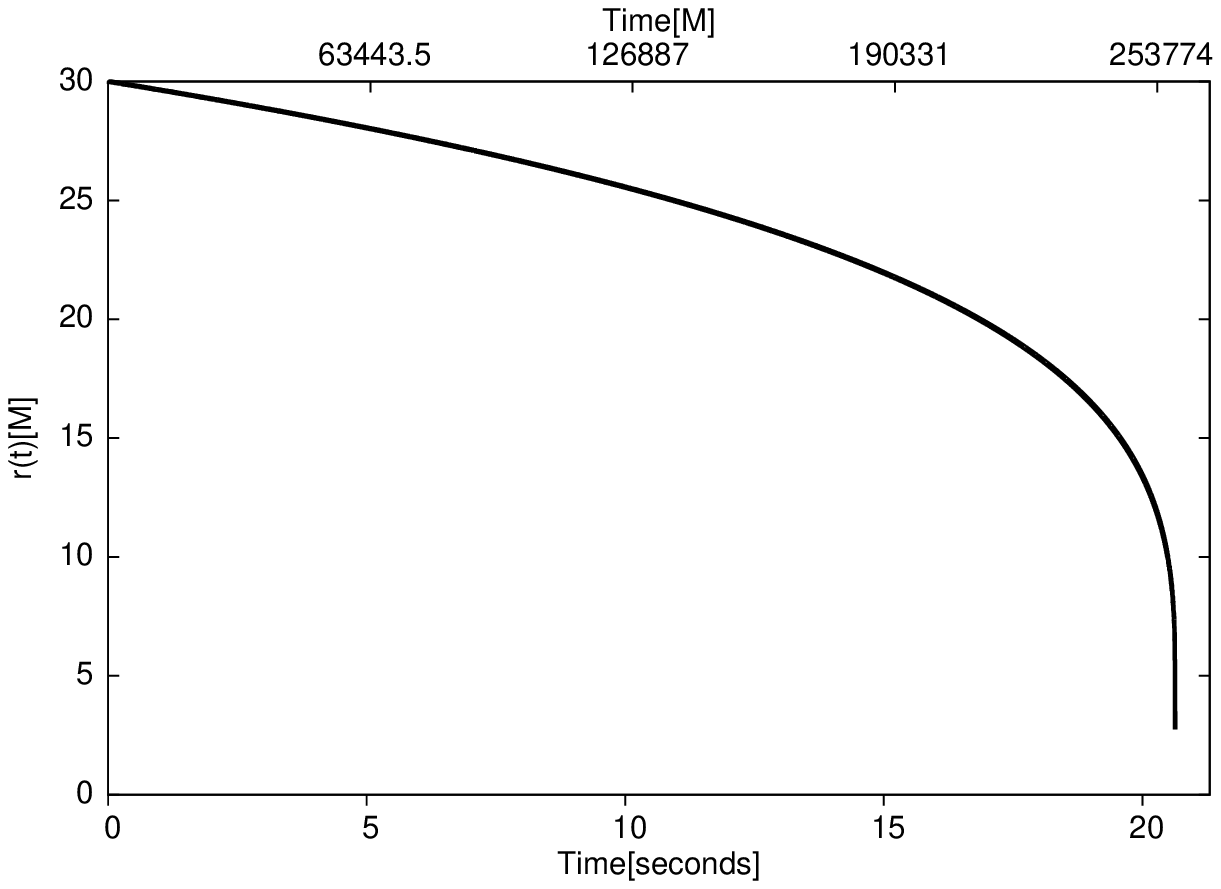}
\includegraphics[height=0.36\textwidth,  clip]{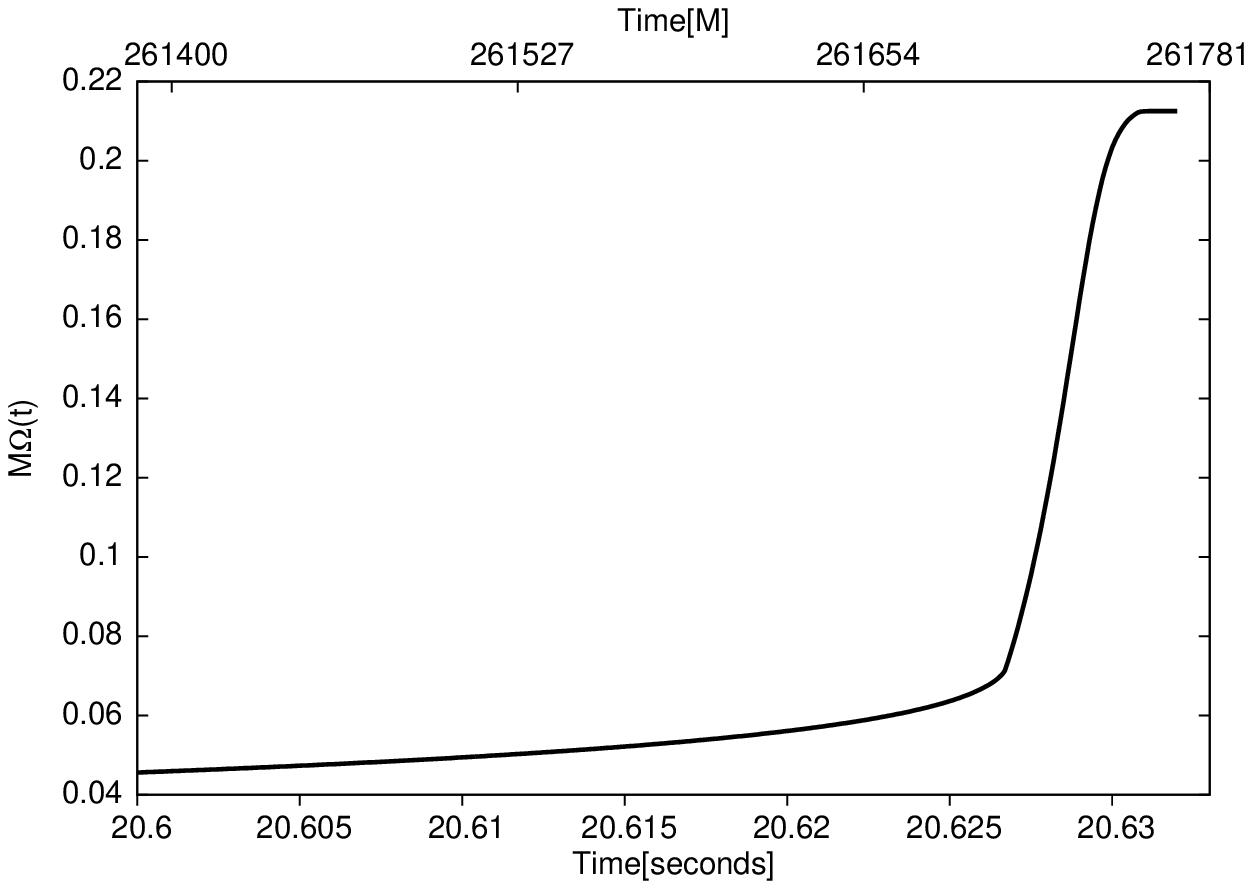}
}
\caption{As in Figure~\ref{IP}, but with the (top, bottom) panels showing the radial and orbital frequency evolution for binaries with mass--ratios \(q=(1/10,\,1/15)\), and total mass \(M\in( 11M_{\odot},\, 16M_{\odot})  \), respectively.  As before, the evolution starts from an initial radial value \(r=30M\).}
\label{IPI}
\end{figure*}

In Figures~\ref{IP} and \ref{IPI} we show the evolution obtained by combining Ori and Thorne's~\cite{amos} transition approach for the radial motion with the frequency evolution proposed by Baker et al~\cite{Baker:2008}.   In all the cases shown in Figures~\ref{IP} and \ref{IPI}, the orbital frequency peaks and saturates at a value given by \(\omega_{\ell m n}/m\). This can be understood if we analyze the asymptotic behavior of Eq.~\eqref{late_frequency} near the light-ring, i.e.,
\begin{equation}
\frac{d \phi}{d \mathrm{t} }  \approx \Omega_{\rm{f}} -  \left(\Omega_{\rm{f}}\  -\Omega_{\rm{i}}\right)e^{ -2(t - t_0)/b}.
\label{frequency_LR}
\end{equation}
\noindent Recasting Eq.~\eqref{late_frequency} in this form, enables us to identify the constant coefficient \(b\) with the e-folding rate for the decay of the fundamental quasinormal mode (QNM).  As discussed previously,  Eq.~\eqref{late_frequency} predicts the expected orbital evolution during late inspiral and onward. To provide a unified description from late inspiral through to ringdown, we have decided to adopt the IRS approach, since this framework allows us to smoothly transition from late inspiral  onto the plunge phase, and finally describe the ringdown waveform as a natural consequence of the IRS strain-rate amplitude decay relation \(A^2(t) \propto \Omega \dot{\Omega}\)~\cite{Baker:2008}. In other words, since Eq.~\eqref{frequency_LR} has the correct behavior near the light-ring as predicted by BHPT, the IRS model provides a natural framework to attach the ringdowm waveform at the end of the plunge phase. We will discuss this feature in further detail in the following Section.

\subsection{Ringdown Waveform}
\label{RDwav}

 Numerical relativity simulations have shown that coalescing binary BHs in general relativity lead to the formation of a distorted rotating remnant, which radiates GWs while it settles down into a stationary Kerr BH~\cite{Berti:2006, Berti:2006b}. The GWs emitted during this intermediate phase resemble a ringing bell. Hence, this type of radiation is commonly known in the literature as ringdown radiation, and consists of a superposition of QNMs --- first discovered in numerical studies of the scattering of GWs in the Schwarzschild spacetime by Vishveshwara~\cite{Vish:1970}.   QNMs are damped oscillations whose frequencies and damping times are uniquely determined by the mass and spin of the post-merger Kerr BH.  The frequency \(\hat \omega\) of each QNM has two components: the real part represents the oscillation frequency, and the imaginary part corresponds to the inverse of the damping time:
 \begin{equation}
 \hat \omega = \omega_{\ell m n} - i/\tau_{\ell m n}.
 \label{omega_QNM}
 \end{equation}
\noindent As discussed above, the observables \(\omega_{\ell m n}, \, \tau_{\ell m n}\) are uniquely determined by the final mass, \(M_{\rm{f}}\), and final spin, \(q_{\rm{f}}\), of the post-merger Kerr BH. To determine  \(M_{\rm{f}}\), we use the analytic phenomenological expression for the final mass of the BH remnant that results from the merger of generic binary BHs on circular quasi-orbits introduced in~\cite{Barausse:2012}, namely,
\begin{equation}
\frac{M_{\rm{f}}}{M} = 1- \left(1-\frac{2\sqrt{2}}{3}\right)\eta - 0.543763\, \eta^2.
\label{finalmass}
\end{equation}
\noindent This expression reproduces the expected result in the test mass particle limit, and also reproduces results from currently available numerical relativity simulations~\cite{Barausse:2012,spif}. We determine the final spin of the BH remnant  \(q_{\rm{f}}\) using the fit proposed in~\cite{Bounanno:2007}:
\begin{equation}
q_{\rm{f}} =    \sqrt{12}\, \eta + s_1\,\eta^2 + s_2\,\eta^3,
\label{finspin}
\end{equation}
\noindent with:
\begin{gather}
s_1=-3.454\pm0.132, \qquad  s_1=2.353\pm0.548.
\label{spin_coeff}
\end{gather}
\noindent This prescription is consistent with the numerical relativity simulations described in~\cite{Bounanno:2007,spif}, and reproduces test mass limit predictions. This compact formula is also consistent with the  prescriptions introduced in~\cite{Barus:2009,Rezzolla:2008}. The largest discrepancy between Eq.~\eqref{finspin} and those derived in~\cite{Barus:2009,Rezzolla:2008}  is \(\lesssim2.5\%\) for binaries with \(q\lesssim1/6\). The ringdown waveform is given by~\cite{Berti:2006b, Baker:2008}
\begin{eqnarray}
h(t)&=& -\left(h_{+} - i h_{\times}\right) = \frac{ M_{\rm{f}} }{D} \sum_{\ell m n} {\cal{A}}_{\ell m n}\,e^{-i \left(  \omega_{\ell m n}\, t  + \phi_{\ell m n} \right)} \, e^{-t/\tau_{\ell m n} }\,, 
\label{waverd}
\end{eqnarray}
\noindent where \(  {\cal{A}}_{\ell m n} \) and \( \phi_{\ell m n}\) are constants to be determined by smoothly matching the plunge waveform onto the subsequent ringdown. The ringdown portion of the self-force waveform model constructed in this paper includes the mode \(\ell=m=2\) and the tones \(n=0,\, 1, \, 2\). The approach we follow to attach the leading mode and overtones is the following: 

\begin{itemize}
\item In order to ensure continuity between the plunge and ringdown waveforms, we use the end of the plunge waveform --- Eqs.~\eqref{insppcor}, \eqref{inspccorrected} --- to construct an interpolation function \(F(t)\). The interpolation method used to construct \(F(t)\) is a cubic spline.
\item We match the plunge waveform onto the leading mode  \(\ell=m=2\), \(n=0\)  of the ringdown waveform, Eq.~\eqref{waverd}, at the point where the amplitude of the plunge waveform peaks, \(t_{\rm{max}}\). Attaching the mode requires \(F(t=t_{\rm{max}})\) and \(F'(t=t_{\rm{max}})\) which are computed from the interpolation function. These conditions fix two constants per polarisation.
\item To attach the first overtone,  \(\ell=m=2,\, n=1\),   we insert into Eq.~\eqref{waverd} the constants determined by attaching the leading mode as seeds to compute the amplitude and phase coefficients for the first overtone by enforcing continuity at  \(t_{\rm{max}} + dt\).
\item Finally, we insert into  Eq.~\eqref{waverd} the value of the amplitude and phase coefficients previously determined for the leading mode and first overtone, and determine the four remaining constants by enforcing continuity at \(t_{\rm{max}} + 2\,dt\).
\end{itemize}

\begin{figure*}[ht]
\centerline{
\includegraphics[height=0.35\textwidth,  clip]{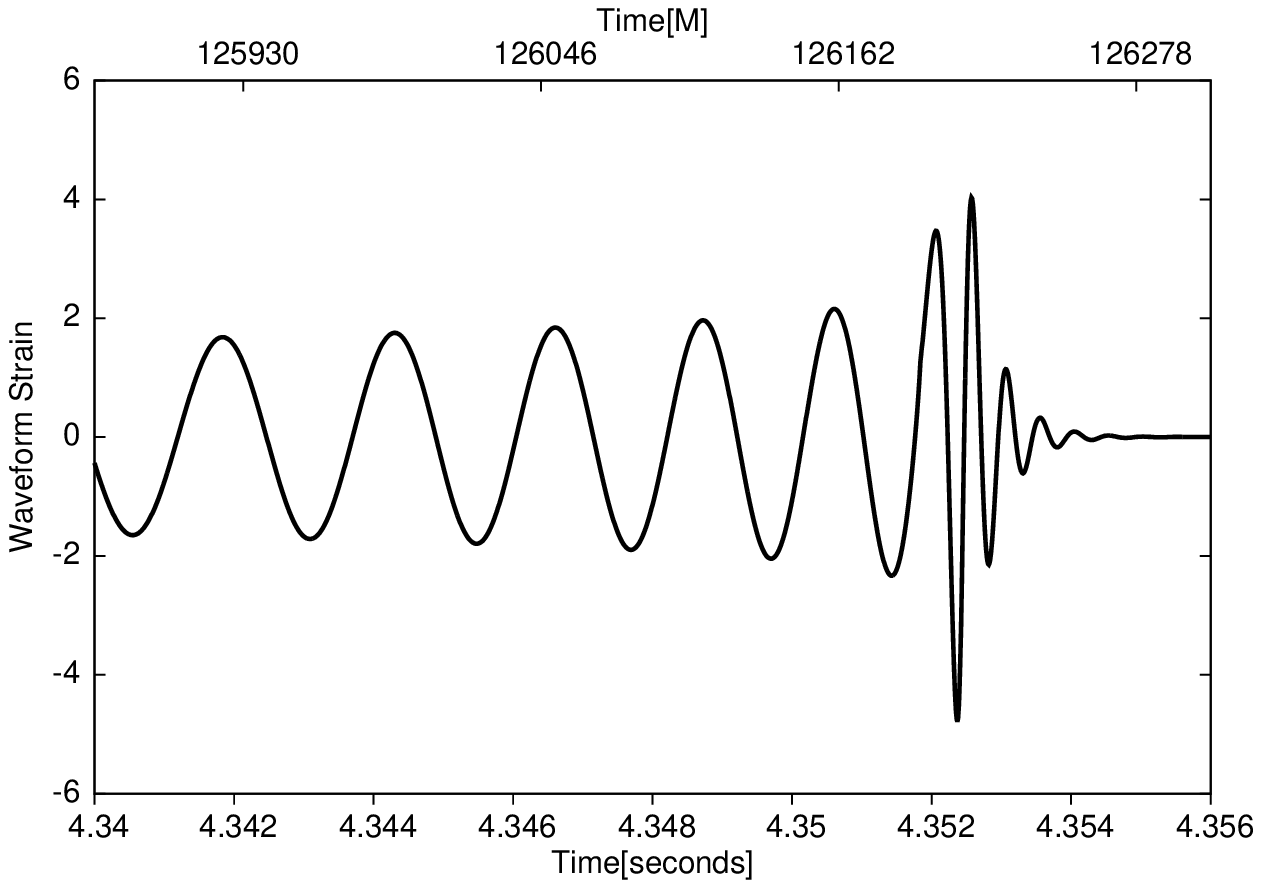}
\includegraphics[height=0.35\textwidth,  clip]{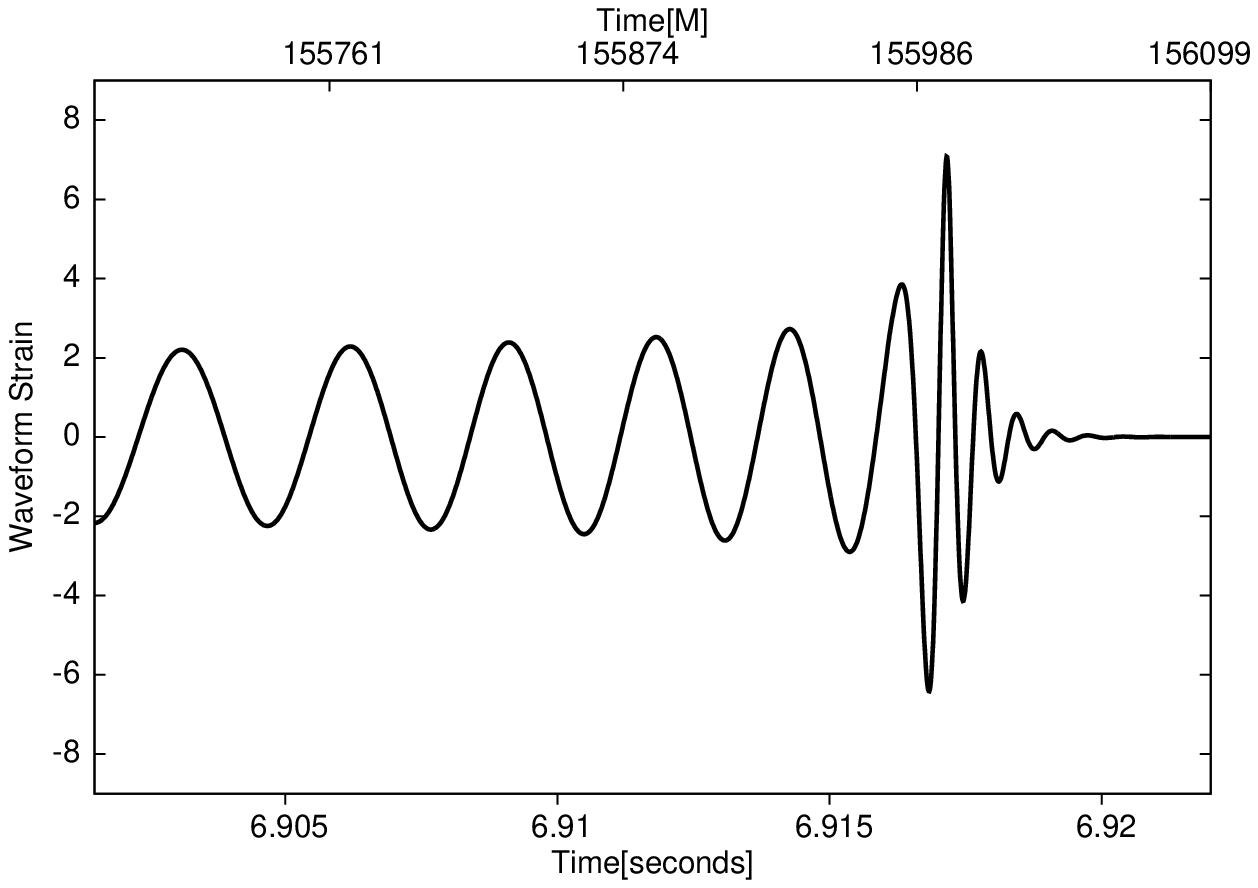}
}
\centerline{
\includegraphics[height=0.35\textwidth,  clip]{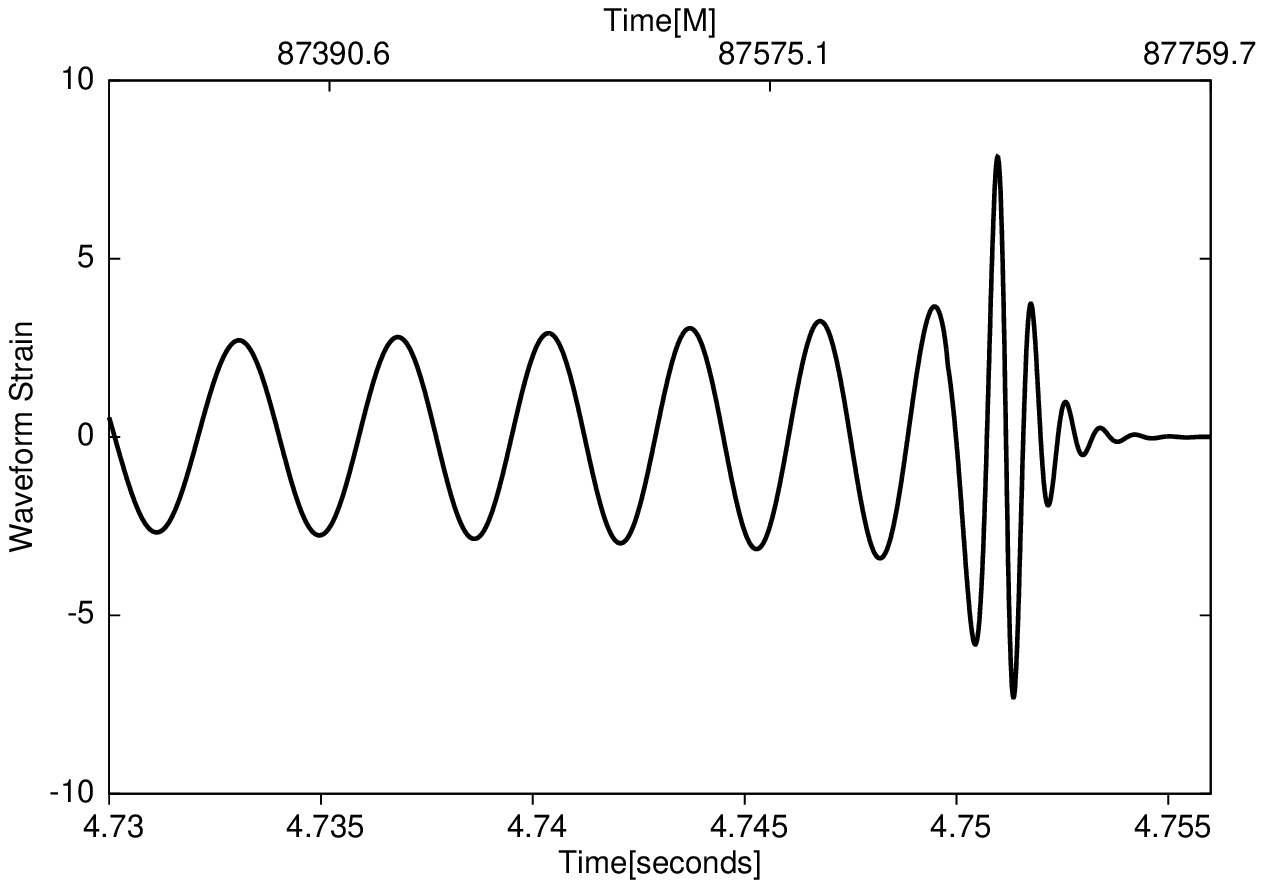}
\includegraphics[height=0.35\textwidth,  clip]{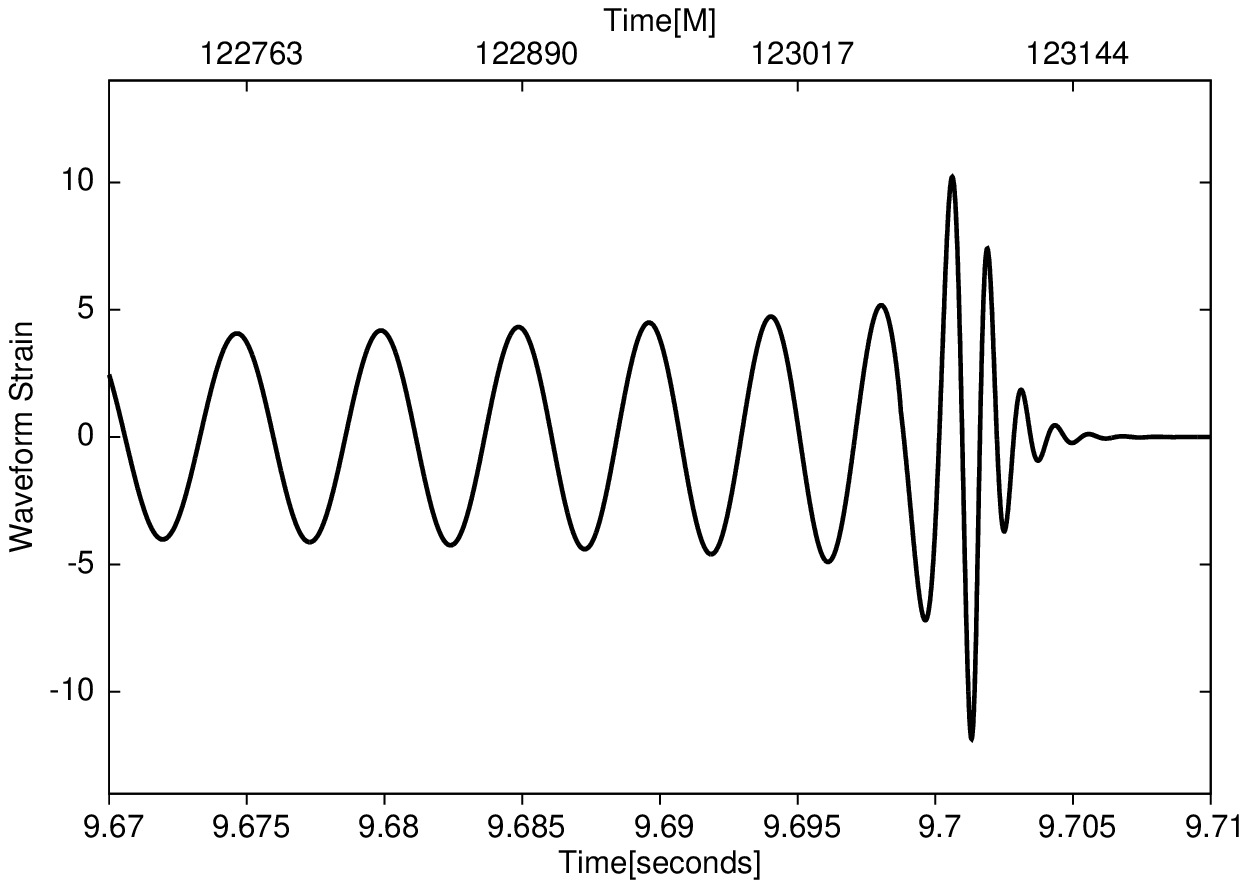}
}
\caption{The panels show sample waveforms from inspiral to ringdown for systems with mass-ratios \(q\in[1/6,\,1/8]\) ---and total mass  \(M\in( 7M_{\odot},\, 9M_{\odot}) \)--- (top panels---from left to right) and \(q\in[1/10,\, 1/15]\)  ---and total mass  \(M\in( 11M_{\odot},\, 16M_{\odot}) \)--- (bottom panels ---from left to right). The inspiral evolution for the [top,\, bottom] panels starts from  \(r=[30M,\, 25M]\). }
\label{Completewavs}
\end{figure*}

Having described the methodology followed to construct complete waveforms for comparable and intermediate mass-ratio systems, we finish this Section by putting together all these various pieces to construct sample waveforms for a few systems with mass-ratio \(q\in[1/6,\,1/8,\, 1/10,\, 1/15]\), and total mass  \(M\in( 7M_{\odot},\, 9M_{\odot},\, 11M_{\odot},\, 16M_{\odot}) \) in Figure~\ref{Completewavs}. 

\subsection{ Ringdown waveform construction in the context of an Implicit Rotating Source}

\begin{figure*}[ht]
\centerline{
\includegraphics[height=0.35\textwidth,  clip]{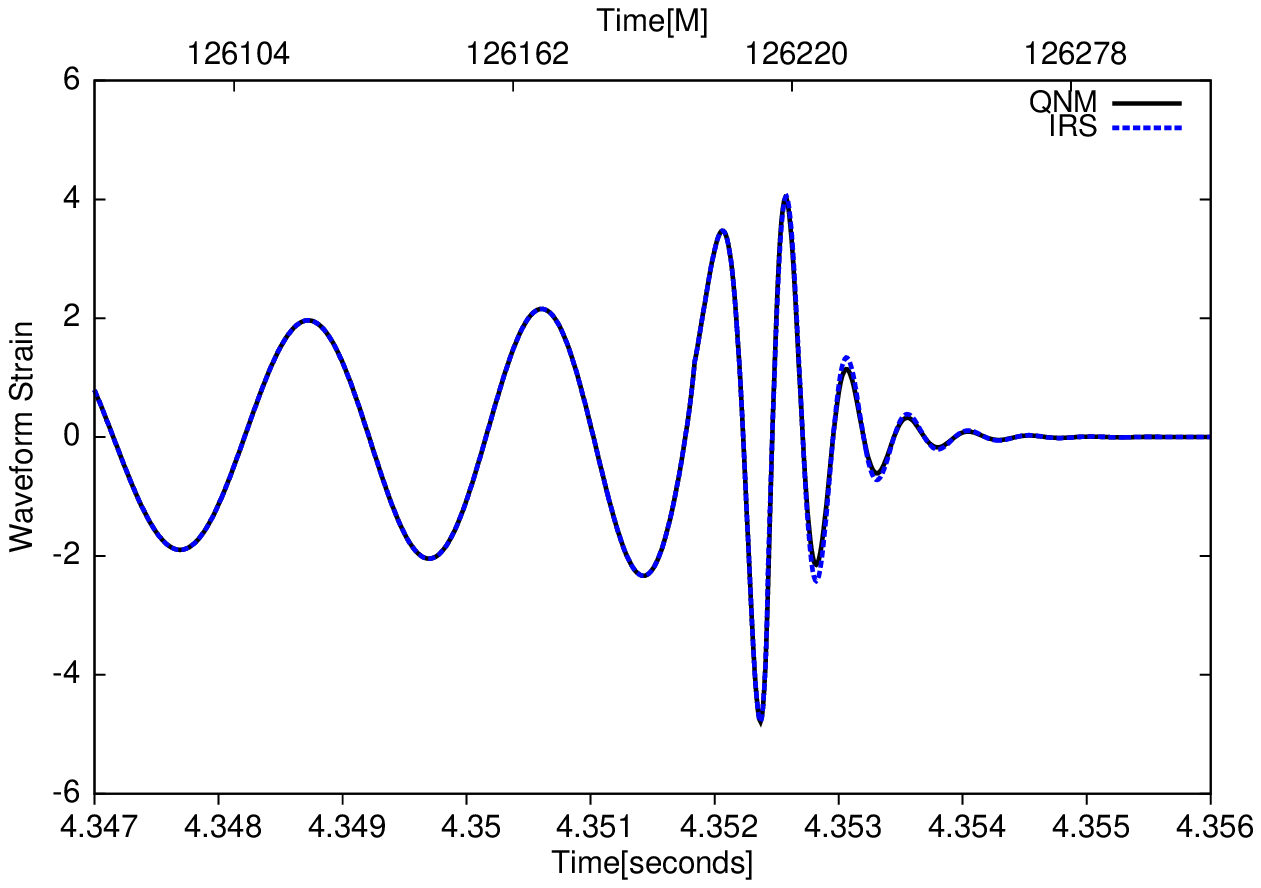}
\includegraphics[height=0.35\textwidth,  clip]{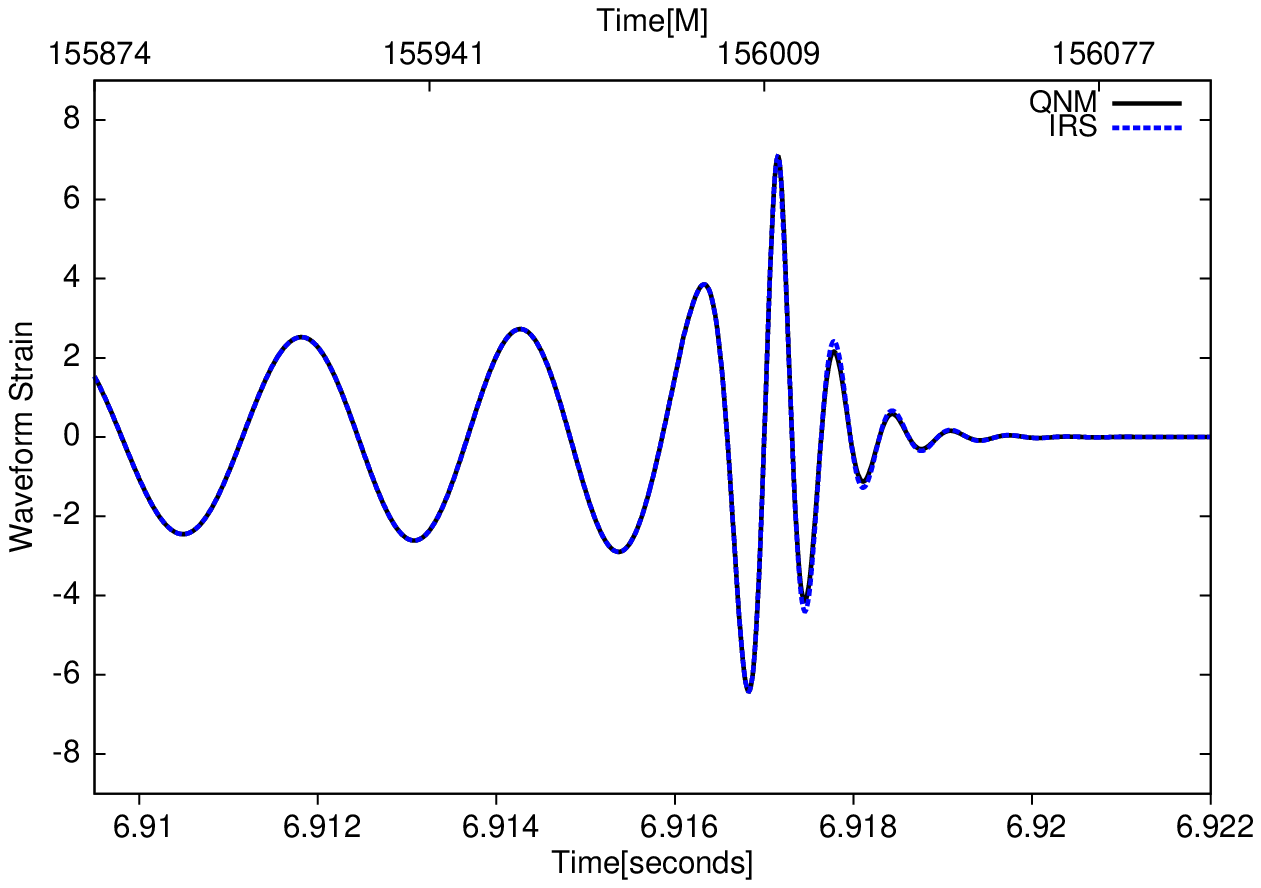}
}
\centerline{
\includegraphics[height=0.35\textwidth,  clip]{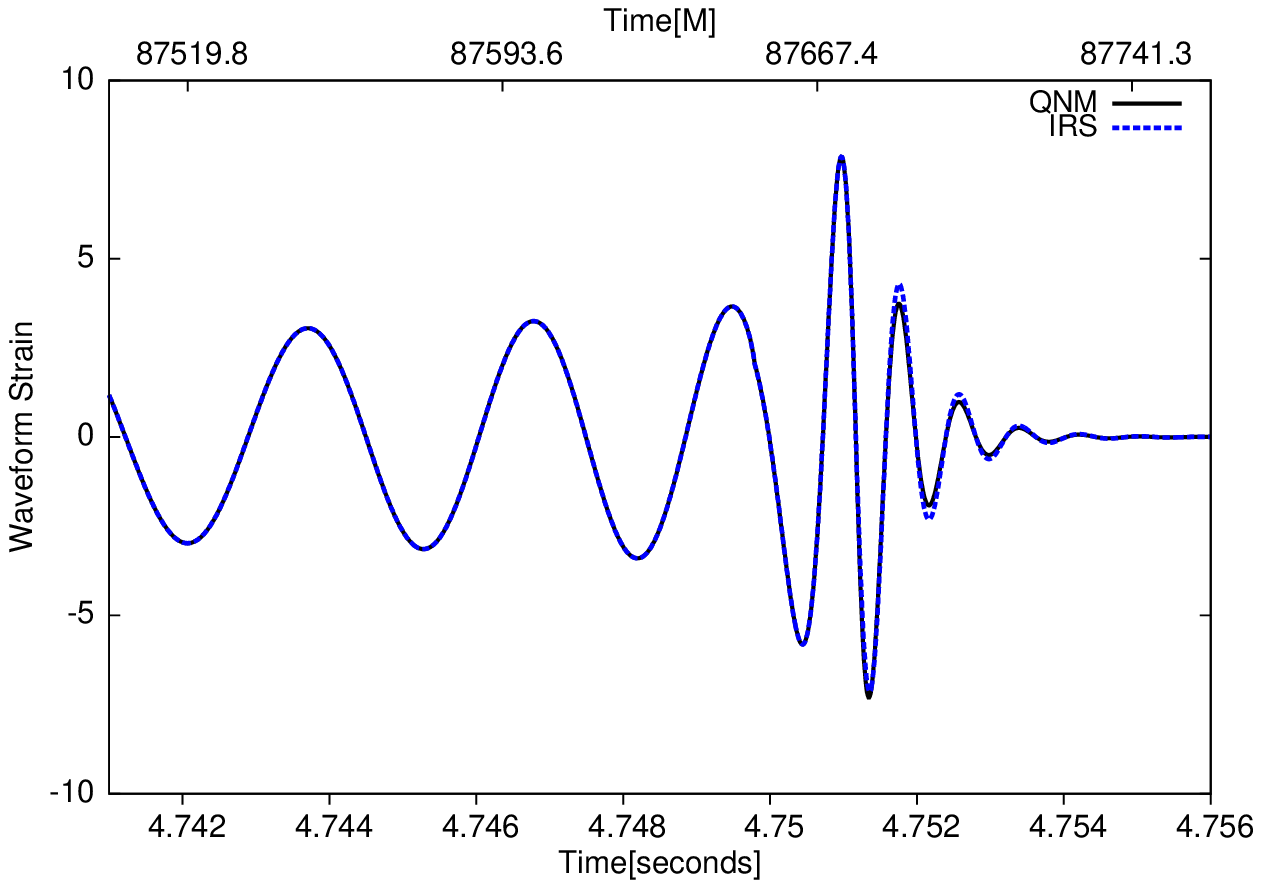}
\includegraphics[height=0.35\textwidth,  clip]{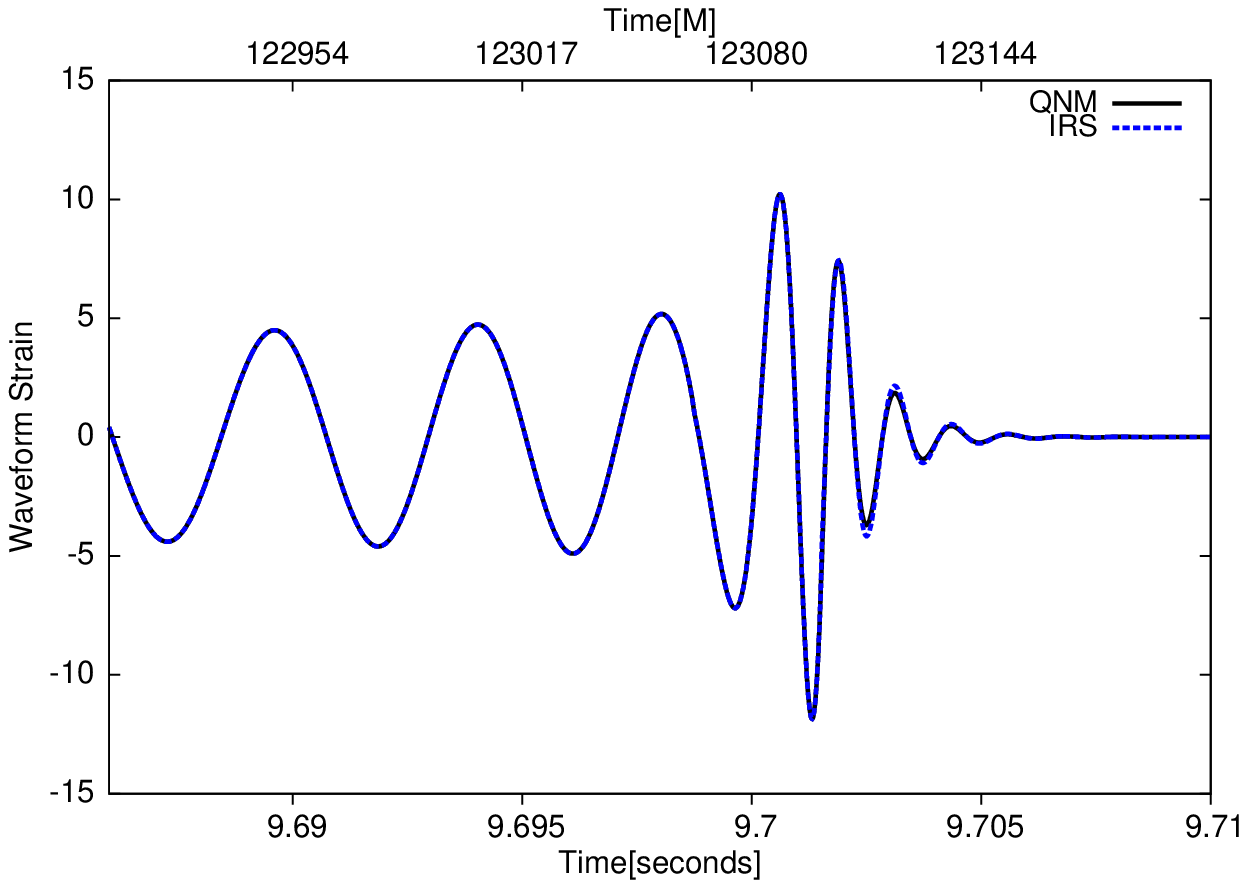}
}
\caption{The panels show sample the late-time evolution of waveforms whose ringdown phase is modeled using the implicit rotating source (IRS) model and a sum of quasinormal modes (QNMs). The systems shown correspond to binaries with mass-ratios \(q\in[1/6,\,1/8]\) ---and total mass  \(M\in( 7M_{\odot},\, 9M_{\odot}) \)--- (top panels---from left to right) and \(q\in[1/10,\, 1/15]\)  ---and total mass  \(M\in( 11M_{\odot},\, 16M_{\odot}) \)--- (bottom panels ---from left to right). The inspiral evolution for the [top,\, bottom] panels starts from  \(r=[30M,\, 25M]\). }
\label{QNMvsIRS}
\end{figure*}

Having described the ringdown waveform construction as a sum of quasinormal modes, we finish this Section  by exhibiting the power of the IRS model to describe the ringdown evolution. In the IRS model, the strain-rate amplitude decay is given by~\cite{Baker:2008}:
\begin{equation}
A^2(t) = 16\,\pi\, \dot{E} \approx 16\,\pi\,\xi\,\Omega\,\dot{\Omega}\,.
\label{amp_rate}
\end{equation}
\noindent Using~Eq.~\eqref{frequency_LR}, in the limit \(\Omega \rightarrow \Omega_{\rm{f}}\), the amplitude decay is given by
\begin{equation}
A_0^2\,e^{-2t/\tau} \approx \frac{32\, \pi\,\xi\, \Omega_{\rm{f}}}{b}\left(\Omega_{\rm{f}} - \Omega_{\rm{i}}\right) e^{-2(t-t_0)/\tau}\,.
\label{rateamp}
\end{equation}
\noindent Hence, the late-time amplitude in the IRS model is given by
\begin{equation}
A^2_{\ell m}  \approx 16\,\pi\,\xi_{\ell m}\,\omega_{\ell m}\,\dot{\omega}_{\ell m}\,,
\label{amp_rate_IRS}
\end{equation}
\noindent where the value of \(\xi_{\ell m}\) is set by ensuring amplitude continuity at the light-ring. In Figure~\ref{QNMvsIRS} we explicitly show the equivalence of the ringdown waveform construction  both in the IRS approach and using the sum of QNMs utilized in the previous Section. This detailed analysis shows that the IRS is a powerful tool  to model the late time portions of the waveforms in a unified way.

\subsection{Summary}
In this Section we briefly summarize the key ingredients that were used to develop the waveform model described in this paper:
\begin{itemize}
\item{Inspiral evolution}
\begin{itemize}
\item The building blocks of the inspiral evolution are the expressions for the energy, \(E\), and angular momentum, \(L\), --- Eqs.~\eqref{enofxeq} and~\eqref{lzofxeq} --- that include gravitational self-force corrections and are valid over the domain \(0<x<\frac{1}{3}\)~\cite{Akcay:2012}.
\item The orbital frequency is modeled using Eq.~\eqref{new_phase}. This prescription encapsulates gravitational self-force corrections that render a better phase evolution when calibrated against EOB.
\item The inspiral trajectory is modeled using the simple prescription~\eqref{radev}. This scheme is no longer valid near ISCO, where \(dE/dx =0\) for binaries with \(q\leq 6\).
\item We construct the inspiral waveform using Eqs.~\eqref{insppcor}, \eqref{inspccorrected}.
\end{itemize}
\end{itemize}

When the inspiralling object nears the ISCO, we need to invoke the `transition scheme' introduced by Ori and Thorne~\cite{amos}, which enables us to smoothly attach the late inspiral evolution onto the plunge phase. 

\begin{itemize}
\item{Transition phase}
\begin{itemize}
\item The transition regime starts at a point when \( \mathrm{d} E/ \mathrm{d}  x\) satisfies Eq.~\eqref{transition_point}.
\item The equations of motion that govern the transition phase are~\eqref{ener_emri}, ~\eqref{ang_emri}. These relations are valid, since the motion near the ISCO is nearly-circular. 
\item Using Eqs.~\eqref{ener_emri}, ~\eqref{ang_emri}, we linearize the second order time derivative of Eq.~\eqref{eomfmrc}.
\item In order to reproduce the orbital phase evolution predicted by numerical simulations from the ISCO to the light-ring, we modify the original transition phase by smoothly matching the inspiral orbital phase evolution, Eq.~\eqref{new_phase}, onto the IRS model, Eq.~\eqref{late_frequency} at the start of the transition phase. 
\end{itemize}
\end{itemize}

\begin{itemize}
\item{Plunge phase}
\begin{itemize}
\item The equations of motion that govern the plunge phase are given by the second order time derivative of Eq.~\eqref{eomfmrc}, and Eq.~\eqref{late_frequency}.
\item We integrate these relations backwards in time to find the point at which both the transition and plunge equations of motion smoothly match. The transition phase ends at this point.
\item Near the light-ring Eq.~\eqref{late_frequency} has the behavior predicted by BHPT, which enables us to smoothly match the plunge phase onto the ringdown. %Put in different words, we don't need to interpolate the orbital frequency evolution to attain the expected value of the orbital frequency at the light-ring. 
\item Both the transition and plunge waveforms are constructed using Eqs.~\eqref{insppcor} and \eqref{inspccorrected}.
\end{itemize}
\end{itemize}

Having constructed the waveform from inspiral to ringdown, we derived second-order radiative corrections to improve the radiative evolution of the waveform model. This was necessary in order to construct a waveform model that is internally consistent, i..e, since the orbital elements include first-order conservative corrections, then radiative corrections should enter the fluxes at second order. We have derived these corrections by enforcing a close agreement between our self-force model and EOB, and have shown that our model can reproduce the orbital phase evolution predicted by EOB within the numerical error of the simulations used to calibrate this model for a variety of mass-ratios. 

\begin{itemize}
\item{Ringdown phase}
\begin{itemize}
\item The ringdown waveform is constructed using Eq.~\eqref{waverd}.
\item We use the plunge waveform to construct an interpolation function \(F(t)\), and then use this function to attach the leading mode \(\ell=m=2,\, n=0\) at the point where the amplitude of the plunge waveform peaks, \(t_{\rm{max}}\). We enforce continuity by ensuring that \(F( t_{\rm{max}}) = h^{n=0}_{\rm{RD}}\) and  \(F'( t_{\rm{max}}) = h'^{n=0}_{\rm{RD}}\).
\item We include the first and second overtone \(n=1,\, n=2\) in the ringdown waveform.
\item  Using the IRS model, we have shown that having knowledge of the time evolution of the orbital frequency provides sufficient information to construct the amplitude decay during ringdowm. Hence, we can construct an alternative ringdowm waveform using only Eq.~\eqref{amp_rate_IRS}, and ensuring smooth continuity with the plunge waveform. 
\item Finally, we have explicitly shown that the implicit rotating source approach provides a natural transition from late-time radiation to ringdown that is equivalent to ringdown waveform modeling based on a sum of QNMs.
\end{itemize}
\end{itemize}

Throughout the paper we have emphasized the fact that our model provides a more reliable framework to model binaries whose components are non-spinning, and with mass-ratios \(q\leq 1/6\), as compared to available PN approximants. It is worth emphasizing that our model is also computationally inexpensive. All the waveforms we generated to constrain the higher-order \(\eta\) corrections in the energy flux ---Eq.~\eqref{etacorrect_new}--- can be generated in fractions of a second. A direct comparison between our code and EOB shows that, averaged over 500 realizations, our code is \(\sim20\%\) faster than the optimized version of the EOB code currently available in the LIGO Scientific Collaboration LAL library. It should be emphasized, though, that our code at present has not been optimized, and hence, compared to EOB our model is expected to significantly reduce the cost of waveform generation, making it relatively more viable for parameter estimation efforts. This is a key feature in our model that enabled us to sample a wide region of parameter space to constrain the \(B_i\) coefficients in Eq.~\eqref{etacorrect_new}. Furthermore, these self-forced waveforms do not need to be highly sampled near the light-ring, hence decreasing the speed with which they can be generated, because the prescription we have used to model the late-time orbital frequency evolution provides the correct evolution near the light-ring. This particular feature also enables us to match the plunge waveform onto its ringdown counterpart without having to interpolate the orbital frequency evolution using a phenomenological approach. The model is internally consistent and the only phenomenology invoked during its construction is related to currently unknown physics, i.e., higher order radiative corrections to the energy flux. Once these corrections are formally derived in the near future, the flexibility of our model will enable us to replace the radiative corrections that we have currently computed by numerical optimization.  At that stage, we will be able to describe in a single unified model the dynamical evolution of binaries whose mass-ratios range from the extreme to the comparable regime. 
%\clearpage 

\section{Conclusions}
\label{conclu}

In this paper we have developed a self-force waveform model that is capable of reproducing with good accuracy the inspiral, merger and ringdown of binaries with mass-ratios \(q\leq 1/6\). This work suggests that a model that incorporates first order conservative self-force corrections in the orbital elements, and second-order radiative corrections in the dissipative piece of the self-force may suffice to describe in a unified manner the coalescence of binaries with mass-ratios that range from the comparable to the extreme. Our model includes conservative self-force corrections that have been derived in the strong--field regime~\cite{Akcay:2012}. Using these results, we find that binaries with mass-ratios \(q\gtrsim1/6\) do not have an ISCO. For systems with \(q\leq1/6\), we have derived a simple relation that provides the location of the ISCO in terms of the symmetric mass-ratio (see Eq.~\eqref{xisco_eq}). To describe the inspiral evolution, we have derived second-order corrections to the energy flux by minimizing the phase discrepancy between our self-force model and the EOB model~\cite{buho, Damour:2013} for a variety of mass-ratios. We have shown that our model reproduces the phase evolution of the EOB model within the accuracy of available numerical simulations for a variety of mass-ratios. 

This paper also presents an extension of the ``transition regime'' developed by Ori and Thorne~\cite{amos} to smoothly match the late inspiral evolution onto the plunge phase. We have found that using the inspiral phase expression for the orbital frequency during the plunge phase does not render an accurate description of the actual orbital evolution as predicted by numerical simulations. Hence, we have embedded the self-force framework in the IRS model proposed by Baker et al~\cite{Baker:2008} to ensure that our model is as close as possible to the orbital evolution predicted by numerical relativity simulations. The implementation of this  prescription ensures that the orbital frequency saturates near the light-ring, which facilitates matching onto the ringdown phase. We have shown that the IRS model provides a natural transition onto the ringdown phase that is equivalent to a ringdown waveform construction based on a sum of QNMs. 

The motivation for constructing this model is two-fold: to exhibit the versatility of the self-force formalism to accurately describe the evolution of binaries beyond the extreme mass-ratio limit; and to provide a tool that can be used to explore the information that could be extracted by GW detectors that target binaries with comparable and intermediate mass-ratios. Current studies have only explored the use of PN approximants to model the merger of NSBH  binaries, despite their inadequacy to capture the evolution of these systems~\cite{Prayush:2013a,  pnbuo, Nitz:2013mxa} (see Figure~\ref{pn_approx}). Comparing  Figures~\ref{pn_approx} and~\ref{PNoptimized}, we conclude that even if second generation GW detectors were only capable of capturing the inspiral evolution of NSBH mergers, our self-force model would be better equipped to describe these events. The construction of this IRS self-force model constitutes an important step towards the construction of more reliable waveforms to describe  IMRCs. In~\cite{Smith:2013}, it was shown that Huerta--Gair (HG) waveforms --- which are closely related to the model developed in this paper, and which were also based on a consistent combination of BHPT and self-force corrections --- work in the relevant mass range for advanced detectors. The new model introduced in this article, improves upon these waveforms and should therefore not only work in the regime of interest to advanced detectors but over a wider parameter space that could be applicable to third generation detectors  or later observations.

Having developed a strong foundation to model binaries on circular orbits whose components are not rotating, it is necessary to incorporate more ingredients to the waveform model to capture GW signals from binaries whose components have significant spin~\cite{Foucart:2012, buho, maeda, burko, smallbody, buoerr1, buoII, TaylorT4Origin}, or systems that form in core-collapsed globular clusters, and hence are expected to have non-negligible eccentricity at merger~\cite{Leary:2009, Huerta:2013a}. In order to do so, we require input both from the self-force program --- which is making substantial progress towards the computation of the self-force in a Kerr background~\cite{Fan:2013b, Sam:2011,Pound:2013}--- and from numerical simulations~\cite{Mroue:2013}, in particular from binaries with typical mass-ratios \(1/20<q<1/10\) to explore the performance of our self-force model in this currently unconstrained regime. Undoubtedly, the development and implementation of improved numerical algorithms~\cite{Fan:2013a}  to carry out these simulations will facilitate the realization of these studies in the near future.

Recent observational discoveries~\cite{Morscher:2013} suggest that NSBH mergers may also occur in globular clusters. Hence, in these dense stellar environments we may expect that multiple \(n\)-body interactions and binary exchange processes may lead to the formation of binaries on eccentric orbits. Detectors such as the Einstein Telescope, operating at low frequencies \(\sim 1\)Hz, may be capable of detecting the signature of eccentricity during the early inspiral. In order to assess these effects, we aim to extend the model introduced in this paper by including eccentricity, making use of self-force corrections for generic orbits in a Schwarzschild background~\cite{wargar}.

%\clearpage

\section*{Acknowledgments}
EH thanks the hospitality of the TAPIR group at Caltech where part of this work was carried out, and the generosity of Prof Peter Saulson, who made this visit possible trough his National Science Foundation (NSF) grant number PHY-120583. We are pleased to thank Chad Galley, Alex Huerta Gago, An{\i}l Zengino\u{g}lu and Fan Zhang for useful interactions. JG's work is supported by the Royal Society. PK acknowledges support from the NSF grant number PHY-0847611. Some calculations were performed on the Syracuse University Gravitation and Relativity cluster, which is supported by NSF grants PHY-1040231, PHY-1104371 and Syracuse University ITS. 

\bibliography{references}

\end{document}